\documentclass[12pt,a4paper]{article}

\usepackage{amsthm,amsmath,amsfonts,amssymb}
\usepackage[authoryear]{natbib}
\usepackage[colorlinks,linkcolor=blue,citecolor=blue,urlcolor=blue]{hyperref}
\usepackage{graphicx}        
\usepackage{bm}            
\usepackage{mathtools}    
\usepackage{pifont}        
\usepackage{booktabs}      
\usepackage{makecell}      
\usepackage{multirow}      
\usepackage{tabularx}
\usepackage{ragged2e}
\usepackage{enumitem}
\usepackage{float}        
\usepackage{algorithm}    
\usepackage{algpseudocode}
\usepackage{authblk}
\usepackage{setspace}
\usepackage[margin=1in]{geometry} 

\def\tcc{\textcolor{blue}}

\theoremstyle{plain}

\newtheorem{theorem}{Theorem}[section]
\newtheorem{lemma}[theorem]{Lemma}
\newtheorem{corollary}[theorem]{Corollary}

\theoremstyle{definition}

\newtheorem{assumption}{Assumption}
\newtheorem{remark}{Remark}

\title{Asymptotic Theory and Sequential Testing for Adaptive Bandits}

\author[1]{Li Yang}

\author[1]{Xiaodong Yan\thanks{Corresponding authors. Email: yanxiaodong@xjtu.edu.cn; jiangdd@xjtu.edu.cn}}

\author[1]{Dandan Jiang\protect\footnotemark[1]}

\affil[1]{School of Mathematics and Statistics, Xi'an Jiaotong University, Xi'an, 710049, China}

\date{}

\begin{document}

\maketitle

\begin{abstract}
        Multi-armed bandit (MAB) processes constitute a foundational subclass of reinforcement learning problems and represent a central topic in statistical decision theory. Yet, conducting valid sequential testing under adaptive allocation remains challenging due to the lack of asymptotic theory under non-i.i.d. reward sequences and sublinear sample sizes for some arms.
        To address this open challenge, we propose an Urn Bandit (UNB) process to integrate the reinforcement mechanism of urn probabilistic models with MAB principles, ensuring almost sure concentration of allocation proportions on optimal arms.
        We establish a joint functional central limit theorem (FCLT) for consistent estimators of expected rewards under non-i.i.d. reward sequences with non-sub-Gaussian tails and pairwise cross-arm dependence.
        To overcome the limitations of existing methods that focus mainly on cumulative regret and therefore provide only algorithmic performance guarantees without supporting valid sequential testing, we develop an asymptotic theory for sequential test statistics under the proposed UNB process. The resulting framework enables a broad class of sequential inference procedures, such as A/B testing and policy evaluation.
        Simulation studies and real data analysis demonstrate that UNB maintains testing performance comparable to that of the equal randomization (ER) design while achieving improved reward accumulation relative to ER.
\end{abstract}

\vspace{1em}

\section{Introduction}
    In the areas of sequential decision making under uncertainty, such as clinical trials, online testing, and policy evaluation, researchers often need to allocate subjects adaptively while conducting sequential testing based on accumulating data \citep{Zhu2010}. Adaptive allocation allows for a more efficient and ethically sound use of resources by preferentially assigning subjects to treatments or arms that perform better \citep{Wei1978,Hu2006}, while sequential testing enables timely inference and early stopping once sufficient evidence has been accumulated \citep{Jennison2000,Gang2021,Wang2024}. The integration of adaptive allocation and sequential testing is fundamental yet challenging for designing efficient experiments that support timely, informed decision making.

    The multi-armed bandit (MAB) framework serves as a statistical foundation for adaptive allocation in sequential experiments.
    By balancing {\it exploration} and {\it exploitation} based on past outcomes, MAB models allocate more units to better performing arms, enhancing the ethicality of adaptive designs \citep{Sutton2018}. The framework has inspired various algorithms, including $\varepsilon$-greedy strategies, upper confidence bound (UCB) methods \citep{Agrawal1995,Auer2002,Garivier2011}, and posterior sampling approaches such as Thompson Sampling \citep{Thompson1933,Agrawal2012,Qi2025}. In practical applications such as recommendation systems, modern MAB extensions incorporate simultaneous arm selection and correlated reward structures to accommodate richer data dependencies \citep{Xia2016,Gupta2021}.

    Efficient adaptive allocation often calls for statistically principled sequential tests to support valid policy learning. The theoretical foundations for sequential tests trace back to \cite{Wald1945}, with subsequent group sequential methods developed for clinical trials to manage interim monitoring, rejection boundary construction, and Type I error control \citep{Pocock1977,Lan1983}. More recently, sequential testing procedures have been widely applied beyond clinical trials, including online multiple testing and false discovery rate control \citep{Gang2021}, online A/B testing \citep{Johari2022,shi2023}, and sparse recovery \citep{Wang2024}.

    Recent research has increasingly focused on integrating adaptive allocation with sequential testing. Specifically, \cite{Zhu2010} justified classical group sequential boundaries for response-adaptive designs where allocation proportions converge to fixed interior targets.
    In contrast, for more aggressive rules like $\varepsilon$-greedy, \cite{Shi2021} employed online bootstrap calibration to address the non-standard covariance structures arising from imbalanced sampling. Alternatively, the always valid inference frameworks leverage $e$-processes and Ville's inequality to achieve finite-sample, time-uniform Type I error control under arbitrary stopping times \citep{Johari2022,ramdas2023gametheoretic}. 
    However, in adaptive bandit environments with sublinear sample sizes for some arms and cross-arm dependence, constructing tractable $e$-processes becomes substantially more challenging since different arms accumulate information at heterogeneous rates, and the resulting dependence structure evolves dynamically. Moreover, recent empirical studies comparing $e$-processes based monitoring with classical group sequential procedures suggest that, under fixed interim analysis schedules, classical group sequential calibration may achieve higher finite horizon detection power \citep{Sokolov2026}. Consequently, restoring a tractable asymptotic joint Gaussian structure remains important for incorporating information time transformations and classical sequential boundary constructions within adaptive allocation frameworks.

    Motivated by challenges arising from sublinear sample sizes and non-i.i.d. reward sequences with cross-arm dependence, we develop an Urn Bandit (UNB) process. The UNB integrates the reinforcement mechanism of probabilistic urn models with multi-armed bandit principles by using multinomial sampling to generate the arm-selection vectors. To accommodate simultaneous adaptive allocation and sequential testing,
    we consider sequential inference on functions of expected rewards under the developed UNB process. Specifically, we consider testing hypotheses of the form
	\begin{equation}\label{eq3}
		H_0: h({\bm\mu}_{[{\mathcal{A}}]})\in {\mathcal{K}}_0\quad {\rm versus}\quad H_1:\ h({\bm\mu}_{[{\mathcal{A}}]})\notin{\mathcal{K}}_0,
	\end{equation}
	where ${\bm \mu}=(\mu_1,\dots,\mu_d)^\top$ denotes the vector of expected rewards, with the subscript $[{\mathcal{A}}]$ highlighting that the samples are collected under the adaptive allocation rule ${\mathcal{A}}=\{{\cal A}_1,{\cal A}_2,\cdots\}$.
	Different choices of the function $h(\cdot)$ accommodate a wide range of practical objectives. For instance, setting $h({\bm\mu}_{[{\mathcal{A}}]})=\mu_1-\mu_2$ corresponds to testing equality between two arm means. Table~\ref{tab1} summarizes the key structural differences between existing approaches and our proposed UNB framework. 
	The key contributions are listed as follows:

    (i) To overcome the limitations of existing MAB methods that primarily evaluate adaptive allocation through cumulative regret analysis, we develop a unified framework that integrates adaptive allocation and sequential testing simultaneously.

	(ii) For the methodology, we propose a new and general MAB process termed UNB, which accommodates non-i.i.d., non-sub-Gaussian rewards with sublinear sample size accumulation for suboptimal arms and pairwise cross-arm correlations. UNB integrates the adaptive reinforcement of the probabilistic urn model with the {\it exploration-exploitation} principle of MAB, implementing an allocation driven by cumulative return via multinomial sampling.
	
	(iii) Theoretically, we establish a functional central limit theorem (FCLT) for the expected reward estimator process under adaptive allocation. Since the cumulative sample size of suboptimal arms grows sublinearly, we apply an information time transformation, after which the sequential test statistic converges to a standard Brownian motion, justifying the use of classical group sequential boundaries.

	The remainder of this paper is organized as follows. In Section \ref{sec2}, we present the UNB process for adaptive allocation. Section \ref{sec3} provides the asymptotic theory and statistical inference. Section \ref{sec4} develops the sequential tests framework within the UNB process. Sections \ref{sec5} and \ref{sec6} evaluate the performance of the proposed method by conducting extensive simulation studies and analyzing real data. Section \ref{sec7} provides concluding remarks.
	Detailed theoretical proofs, additional numerical simulations, and real data studies are provided in the Supplementary Material.

		\begin{table}[h]
		\centering
		\caption{Comparison of adaptive designs across key aspects: adaptivity of allocation, dependence across arms, non-sub-Gaussian rewards, sublinear information growth (i.e., sublinear sample size accumulation for suboptimal arms), FCLT for reward estimators, and support for sequential tests.} 
		\label{tab1} 
		\footnotesize
		\scalebox{0.8}{
			\resizebox{\linewidth}{!}{
				\begin{tabular}{lcccccc}
					\toprule
					& \makecell{Adaptivity\\Allocation}
					& \makecell{Dependence\\across arms}
					& \makecell{Non-sub\\Gaussian}
					& \makecell{Sublinear\\Information}
					& FCLT
					& \makecell{Sequential\\Test} \\
					\midrule
                    \cite{May2009} & \ding{51} & \ding{55} & \ding{51} & \ding{51} & \ding{55} & \ding{55} \\
					\cite{Zhu2010} & \ding{51} & \ding{55} & \ding{51} & \ding{55} & \ding{51} & \ding{51} \\
					\cite{Zhu2012} & \ding{51} & \ding{55} & \ding{51} & \ding{55} & \ding{51} & \ding{51} \\
                    \cite{Gaharwar2020} & \ding{51} & \ding{51} & \ding{55} & \ding{51} & \ding{55} & \ding{55} \\
					\cite{Shi2021} & \ding{51} & \ding{55} & \ding{55} & \ding{55} & \ding{51} & \ding{51} \\
					\cite{Qi2025} & \ding{51} & \ding{55} & \ding{55} & \ding{51} & \ding{55} & \ding{55} \\
					\cite{Chen2025} & \ding{51} & \ding{55} & \ding{55} & \ding{51} & \ding{55} & \ding{55} \\
					\textbf{UNB (Ours)} & \ding{51} & \ding{51} & \ding{51} & \ding{51} & \ding{51} & \ding{51} \\
					\bottomrule
				\end{tabular}
		}}
	\end{table}

	\section{Urn Bandit Process and Test Problem}\label{sec2}
	\subsection{The UNB Process for Adaptive Allocation}
        Consider a $d$-armed bandit problem where an agent sequentially interacts with the environment. The agent's decision making is governed by a novel UNB policy. Let $[1:d]:=\{1,2,\dots,d\}$ be the set of arms. At each stage $n$, the agent performs a sequence of $N_n$ actions, which is adaptively generated based on historical observations. Specifically, the action sequence at stage $n$ is defined as
        ${\cal A}_n=\{a_{n1},a_{n2},\dots,a_{n,N_n}\}$ with $a_{nq}\in [1:d]$, where selections of the same arm within a round are allowed. For arm $k\in[1:d]$, let
        \begin{equation*}
            X_{nk}=\sum_{s=1}^{N_n}\mathbb{I}\{a_{ns}=k\}
        \end{equation*}
        denote the number of times arm $k$ is selected at stage $n$. Then, $\sum_{k=1}^d X_{nk}=N_n$.

        To capture cross-arm dependence, we introduce a latent joint reward process. At each stage $n$, let $\{\bm{\xi}_n^{(q)}\}_{q \ge 1}$ be a sequence of $d$-dimensional random vectors, where $\bm{\xi}_n^{(q)}=(\xi_{n1,q}, \dots, \xi_{nd,q})^\top$ collects the potential rewards for all $d$ arms under a shared latent environment, thereby inducing dependence across arms. For each arm $k$, the observed rewards correspond to the first $X_{nk}$ elements of its latent sequence, given by $\{\xi_{nk,q}\}_{q=1}^{X_{nk}}$. This construction preserves the underlying cross-arm dependence structure under random and unequal sample sizes.

        For each arm $k$, the cumulative return at stage $n$ is updated by
        \begin{equation*}
            R_{nk}=R_{n-1,k}+\sum_{q=1}^{X_{nk}}\xi_{nk,q},
        \end{equation*}
        with the initial value $R_{0k}>0$. Let ${\bf R}_n=(R_{n1},R_{n2},\dots,R_{nd})^\top$ be the cumulative return vector. Define the normalized return vector ${\bf Z}_n=(Z_{n1},Z_{n2},\dots,Z_{nd})^\top$ as
        \begin{equation*}
            {\bf Z}_n=\frac{{\bf R}_n}{\|{\bf R}_n\|_1},
        \end{equation*}
        where $\|\cdot\|_1$ denotes the $\ell_1$-norm.
        
        We now specify the reinforcement driven mechanism of the UNB process. Let $\mathcal{F}_0=\varnothing$ and $\mathcal{F}_n=\sigma({\bf X}_i, \{{\bm\xi}^{(q)}_{i}\}_{q\ge1}:1\le i\le n)$ and ${\cal G}_n=\sigma({\cal F}_n,{\bf X}_{n+1})$. Conditioned on ${\cal F}_{n-1}$, the arm selection count vector ${\bf X}_n=(X_{n1},X_{n2},\dots,X_{nd})$ follows a multinomial distribution,
        \begin{equation*}
            {\bf X}_n\sim {\rm Multinomial}(N_n,{\bf Z}_{n-1}).
        \end{equation*}
        This mechanism induces a reinforcement structure that assigns higher selection probabilities to arms with larger cumulative return, while maintaining persistent exploration. Algorithm \ref{alg:unb_process} provides a formal description.

    \begin{algorithm}[H]
    \caption{The UNB Process for Adaptive Allocation}
        \label{alg:unb_process}
        \footnotesize
        \begin{minipage}{\linewidth}
        \begin{algorithmic}[1]

            \Require Horizon $T$, burn-in period $n_0$, and batch sizes $\{N_n\}_{n=1}^T$.

            \State \textbf{Initialization:} Pull each arm $n_0$ times, observe the corresponding rewards, and initialize the cumulative return vector ${\bf R}_0$.

        \For{$n=1$ to $T$}

            \State Generate allocation vector ${\bf X}_n \sim \mathrm{Multinomial}(N_n; {\bf Z}_{n-1})$.

            \State For each arm $k\in [1:d]$, select arm $k$ exactly $X_{nk}$ times, forming the action sequence ${\cal A}_n$.

            \State Observe rewards $\{\xi_{nk,q}:1\le q \le X_{nk},\ k\in [1:d]\}$.

            \State Update cumulative return:
                    $R_{nk}=R_{n-1,k} + \sum_{q=1}^{X_{nk}} \xi_{nk,q}, \quad k \in [1:d]$.
        
            \State Update normalized return vector:
                    ${\bf Z}_n=\frac{{\bf R}_n}{\|{\bf R}_n\|_1}$.
        \EndFor
        \end{algorithmic}
        \end{minipage}
        \end{algorithm}

    We assume that the total budgets $\{N_n\}_{n\ge1}$ form a bounded sequence of positive integers, and converge to $N$ as $n\to\infty$.
    The reward process is independent across stages $n$ and not necessarily identically distributed. For each stage $n$, the sequence $\{\bm{\xi}^{(q)}_n\}_{q\ge1}$ is i.i.d. in $q$.
    For each arm $k$, we assume that the mean and variance of the reward distribution at stage $n$ satisfy $\mu_{k,n}=\mathbb{E}(\xi_{nk,q})\to\mu_k$ and $\sigma^2_{k,n}={\mathbb V}{\rm ar}(\xi_{nk,q})\to \sigma^2_k$. Furthermore, we assume that for any $k \neq s$, the covariance and correlation satisfy
    $C_{ks,n}=\mathbb{C}{\rm ov}(\xi_{nk,q}, \xi_{ns,q}) \to C_{ks}$ and $\rho_{ks,n}=\mathbb{C}{\rm orr}(\xi_{nk,q},\xi_{ns,q})\to\rho_{ks}$ as $n\to\infty$.

    To characterize the long-term behavior of the UNB process, let 
    \begin{equation*}
        \mu^\ast=\max\{\mu_1,\mu_2,\dots,\mu_d\},\qquad  \mu_\ast=\min\{\mu_1,\mu_2,\dots,\mu_d\},
    \end{equation*}
    and let ${\mathcal{I}}=\{k:\mu_k=\mu^\ast\}$ denote the set of optimal arms. Define the cumulative sample size $S_{nk}$ for arm $k$ and the cumulative total sample size $S_n$ up to stage $n$ as
	\begin{align}\label{eqrr2}
		S_{nk}=\sum_{t=1}^nX_{tk}, \qquad S_n=\sum\limits_{k=1}^dS_{nk},
	\end{align}
    respectively. We now impose the following assumptions on the reward process to facilitate the asymptotic analysis of the allocation dynamics under the UNB framework.

    \begin{assumption}\label{meanbound}
        There exists a constant $C$ such that $\sup _{n,k,q} \mathbb{E}\big(\xi^3_{n k,q}\big)<C$.
    \end{assumption}
    
	\begin{assumption}\label{meanconver_weak}
        For all $k\in[1:d]$, $|\mu_{k,n}-\mu_k|=O(n^{-\epsilon})$ for some $\epsilon>0$.
    \end{assumption}
    
    \begin{assumption}\label{meanconver}
        For all $k\in[1:d]$, $|\mu_{k,n}-\mu_k|=o\big(n^{-\frac{\mu_k}{2\mu^\ast}}\big)$.
    \end{assumption}

     Assumption \ref{meanbound} requires only uniformly bounded third moments, allowing reward distributions beyond the sub-Gaussian class. 
     Assumptions \ref{meanconver_weak} and \ref{meanconver} relax the classical time-homogeneity assumption on expected rewards by allowing time-varying means that converge at specified rates. Assumption \ref{meanconver_weak} is used for the limit theory of the UNB allocation, while Assumption \ref{meanconver} is required for the CLT of the estimators of $\mu_k$ and subsequent inference.

     The first result establishes the strong consistency of the normalized return vector $\mathbf{Z}_n$. Specifically, it shows that the self-reinforcing mechanism leads to asymptotically negligible contributions from suboptimal arms to the total return.

     \begin{theorem}\label{thr1}
        Suppose that Assumptions \ref{meanbound} and \ref{meanconver_weak} hold. Then, for each arm $k\in[1:d]$, there exists a random variable $Z_k$ such that
        \begin{equation*}
            Z_{nk}\xrightarrow{{\rm a.s.}} Z_k \quad \text{as }  n \to \infty.
        \end{equation*}
        Furthermore, for any suboptimal arm $k$ with $\mu_k< \mu^\ast$, $Z_k=0$ almost surely.
	\end{theorem}

    As a direct consequence of the limiting allocation proportions, we obtain law-of-large-numbers type limits for both cumulative return and sample sizes.

    \begin{corollary}\label{cor1}
       Suppose that Assumptions \ref{meanbound} and \ref{meanconver_weak} hold. Then, for each arm $k\in[1:d]$,
        \begin{align*}
            \frac{R_{nk}}{n}\xrightarrow{{\rm a.s.}}
                N\mu_k Z_k\quad\text{and}\quad \frac{S_{nk}}{n}\xrightarrow{{\rm a.s.}}NZ_k\quad \text{as } n\to \infty,
        \end{align*}
        where $Z_k=0$ almost surely whenever $\mu_k<\mu^\ast$. Moreover,
        \begin{equation*}
         \frac{\|{\bf R}_n\|_1}{n}\xrightarrow{{\rm a.s.}}N\mu^\ast\quad  \text{and}\quad \frac{\|{\bf S}_n\|_1}{n}\xrightarrow{{\rm a.s.}}N\quad\text{as } n\to \infty.
        \end{equation*}
    \end{corollary}

    While the preceding results establish that the limiting return mass concentrates on the set of optimal arms, Theorem \ref{thrr1} further characterizes the structure of this limiting allocation within the optimal set, distinguishing between the unique and multiple optimal arm cases.
    
    \begin{theorem}\label{thrr1}
        Suppose that Assumptions \ref{meanbound} and \ref{meanconver_weak} hold. Then, if there exists a unique $k^\ast\in[1:d]$ such that $\mu_{k^\ast}=\mu^\ast$, then $Z_{k^\ast}=1$ holds almost surely. Otherwise, for each $k$ such that $\mu_k=\mu^\ast$, it holds that $\mathbb{P}(Z_k\in(0,1))=1$.
    \end{theorem}

    While Theorem \ref{thr1} establishes the asymptotic vanishing of suboptimal arms, the following result provides a refined quantitative analysis by establishing the exact sublinear accumulation rates for their sample sizes and return.

    \begin{theorem}\label{thr6}
        Suppose that Assumptions \ref{meanbound} and \ref{meanconver_weak} hold. Then, for each suboptimal arm $k$ with $\mu_k<\mu^\ast$, there exists an almost surely finite random variable $\tilde{Z}_k$ such that
        \begin{equation*}
            n^{1-\frac{\mu_k}{\mu^\ast}}Z_{nk}\xrightarrow{{\rm a.s.}}\tilde{Z}_k\quad \text{as } n \to \infty.
        \end{equation*}
        Moreover, the cumulative return and selection counts satisfy
        \begin{equation*}
            \frac{R_{nk}}{n^{\mu_k/\mu^\ast}}\xrightarrow{{\rm a.s.}}\mu^\ast N\tilde{Z}_k\quad\text{and}\quad             \frac{S_{nk}}{n^{\mu_k/\mu^\ast}}\xrightarrow{{\rm a.s.}}
                \frac{\mu^\ast}{\mu_k}N\tilde{Z}_k\quad \text{as } n \to \infty.
        \end{equation*}
    \end{theorem}

        The UNB policy exhibits a self-reinforcing mechanism that achieves optimal arm identification. Specifically, Theorems \ref{thr1}, \ref{thrr1} and Corollary \ref{cor1} show that the asymptotic return proportions of suboptimal arms vanish almost surely, whereas optimal arms retain positive limiting proportions. Furthermore, Theorem \ref{thr6} shows that for any suboptimal arm $k$, its sample size satisfies $S_{nk} = O_{\mathrm{a.s.}}(n^{\mu_k/\mu^\ast})$, which implies that $S_{nk}$ grows sublinearly in $n$, ensuring sufficient exploration while preserving allocation efficiency. These established properties provide the foundation for the subsequent asymptotic theory.

        Classical bandit algorithms such as $\varepsilon$-greedy and UCB are primarily designed for fixed exploration policies or regret minimization objectives. Their allocation rules do not generally yield a tractable distributional structure for sequential test statistics, so valid inference typically requires additional calibration. In contrast, the UNB process provides a unified framework for joint allocation and inference. Its reinforcement mechanism based on cumulative return yields a tractable probabilistic structure that admits asymptotic analysis. Moreover, optimal arms are sampled at a linear order while suboptimal arms are explored at a sublinear rate. This guarantees sufficient information accumulation for sequential inference, thereby enabling valid hypothesis testing under adaptive allocation.

	\subsection{Hypothesis on Bandit's Arms}\label{sec2.2}
	
	Our primary inferential goal is to test hypotheses concerning some functionals of the expected rewards from the $d$ arms, ${\bm\mu}=(\mu_1,\mu_2,\dots,\mu_d)^\top$, by adopting the testing framework (\ref{eq3}). We highlight two representative instances.
	
	{\bf a) Linear Combinations of Arms.} We consider linear hypotheses of the form:
	\begin{equation}\label{eq33}
		H_0:\ {\bm \beta}^\top{\bm\mu}_{[{\mathcal{A}}]}\in \mathcal{K}_0\ \ \ {\rm versus}\ \ \ H_1:\ {\bm \beta}^\top{\bm\mu}_{[{\mathcal{A}}]}\notin \mathcal{K}_0,
	\end{equation}
	where ${\mathcal{K}}_0$ is a prespecified constraint set, and ${\bm\beta}\in\mathbb R^d$ specifies the contrast of interest. 
	Different choices of ${\bm\beta}$ and ${\mathcal{K}}_0$ recover a broad range of familiar testing problems. Let ${\bf e}_k$ denote the $k$th standard basis vector in $\mathbb{R}^d$. Specifically:
	
	\textbullet \ {\it A/B Testing}. Setting ${\bm\beta}={\bf e}_i-{\bf e}_j$ and ${\mathcal{K}}_0=(-\infty,0]$ yields $H_0:\mu_i-\mu_j\le0$, representing typical A/B testing \citep{johari2017peeking,zhang2025strategic}.
	
	\textbullet \ {\it Benchmarking against a Threshold}.  Choosing ${\bm\beta}={\bf e}_k$ and ${\mathcal{K}}_0=(-\infty,K_0]$ reduces the problem to assessing whether the expected reward of a single arm does not exceed a fixed benchmark $K_0$ \citep{locatelli2016,kano2019}.
	
	\textbullet \ {\it Comparison with Control Group Average}. More structured comparisons, such as evaluating a new arm against the average of multiple controls, are obtained by contrasts such as ${\bm\beta}=(1,-0.5,-0.5,0,\dots,0)^\top$, typically with ${\mathcal{K}}_0=(-\infty,0]$.
	
	\textbullet \ {\it General Weighted Objectives}. The framework \eqref{eq33} also covers general inference on weighted means by taking ${\bm\beta}={\bm w}$ for any prespecified weight vector ${\bm w}$.

	{\bf b) Nonlinear Functionals of Arms.}
	Beyond linear contrasts, the framework also accommodates nonlinear functionals of ${\bm\mu}$. Let $h:{\mathbb R}^d\to{\mathbb R}$ and consider 
	\begin{equation}\label{eq34} H_0:h({\bm\mu}_{[{\mathcal{A}}]})\le0\quad{\rm versus}\quad H_1:h({\bm\mu}_{[{\mathcal{A}}]})>0, 
	\end{equation} 
	A representative example is inference for relative effects, such as $h({\bm\mu}_{[{\mathcal{A}}]})=\mu_i/\mu_j$, which corresponds to relative lift measures commonly used in online controlled experiments\citep{Larsen2024}. 

    While these inference problems are well-studied under static designs, UNB changes the inferential context by generating adaptive and history-dependent data. Such integration is attractive in multi-arm experiments where both ethical considerations and statistical efficiency favor making timely decisions.

	\section{Asymptotic Theory and Statistical Inference}\label{sec3} 
    This section develops the asymptotic theory under the UNB allocation, introduces the test statistic for problem \eqref{eq3}, and establishes its asymptotic power properties.

	\subsection{Joint Asymptotic Distribution of the Mean Estimators}
	For each arm $k$, a natural estimator for the expected reward $\mu_k$ is the empirical mean of the observed samples given by
    \begin{equation*}
        \hat\mu_{k,n}=\frac{R_{nk}}{S_{nk}}=\frac{R_{0k}+\sum_{t=1}^n\sum_{q=1}^{X_{tk}}\xi_{tk,q}}{\sum_{t=1}^nX_{tk}},
    \end{equation*}
    where the sample size $S_{nk}$ is defined in \eqref{eqrr2}. The strong consistency of this estimator is established in Lemma \ref{ler1}. Under the UNB allocation, the sample sizes are random and history-dependent, leading to a nonstandard asymptotic distribution of the estimators $\hat\mu_{k,n}$ driven by adaptive allocation and cross-arm dependence. The following theorem establishes the joint CLT for $(\hat\mu_{1,n},\dots,\hat\mu_{d,n})^\top$ and characterizes the corresponding covariance structure.

    \begin{theorem}\label{th2}
		Suppose that Assumptions \ref{meanbound} and \ref{meanconver} hold. Then
        \begin{equation}
            \begin{pmatrix}\sqrt{S_{n1}}(\hat\mu_{1,n}-\mu_1),\sqrt{S_{n2}}(\hat\mu_{2,n}-\mu_2),\dots,\sqrt{S_{nd}}(\hat\mu_{d,n}-\mu_d)\end{pmatrix}^\top
            \xrightarrow{{\rm d(stably)}}{\cal N}({\bf 0},{\bm\Sigma}),
        \end{equation}
        where the covariance matrix $\bm{\Sigma}=(\Sigma_{ks})_{1\le k,s\le d}$ is given as follows. For all $k\in[1:d]$,
        \begin{align*}
			[{\bm{\Sigma}}]_{kk}=
				{\sigma}^2_k,
        \end{align*}
        and for $k\neq s$,
        \begin{align*}
        [{\bm{\Sigma}}]_{ks}=
			\begin{cases}
				{C}_{ks}f(N,Z_{k},Z_{s}), & \text{for } \mu_k=\mu_s=\mu^\ast, \\
				0, & \text{for } \mu_k<\mu^\ast\text{ or }\mu_s<\mu^\ast.
			\end{cases}
        \end{align*}
        Here the function $f$ is defined by
        \begin{equation*}
            f(N,Z_{k},Z_{s})=\frac{1}{N\sqrt{Z_k Z_s}}\mathbb{E}[\min(X,Y)],
        \end{equation*}
        where $(X,Y,N-X-Y)\sim {\rm Multinomial}(N;Z_k,Z_s,1-Z_k-Z_s)$. Equivalently,
        \begin{align*}
            f(N,Z_{k},Z_{s})=\sum_{x=1}^{N} \sum_{y=1}^{N-x} \min(x, y)\frac{(N-1)!}{x!y!(N-x-y)!}Z^{x-\frac{1}{2}}_kZ^{y-\frac{1}{2}}_s(1-Z_k-Z_s)^{N-x-y}.
        \end{align*}
    \end{theorem}

    \begin{remark}
        To illustrate the cross-covariance structure, we provide explicit evaluations of the function $f(N,Z_k,Z_s)$ for several specific values of $N$. Specifically,
        \begin{itemize}
        \item When $N=1$, $f(N,Z_k,Z_s)=0$;
        \item When $N=2$, $f(N,Z_k,Z_s)=\sqrt{Z_kZ_s}$;
        \item When $N=3$, $f(N,Z_k,Z_s)=\sqrt{Z_kZ_s}(Z_k+Z_s)+2\sqrt{Z_kZ_s}(1-Z_k-Z_s)$;
        \item When $N=4$, $f(N,Z_k,Z_s)=\sqrt{Z_kZ_s}[Z^2_s+Z^2_k+3Z_kZ_s+3(1-Z_k-Z_s)]$.
        \end{itemize}      
    \end{remark}

    The convergence in Theorem \ref{th2} (and subsequently Theorem \ref{th1}) is in the sense of \textit{stable convergence} \citep{renyi1963}, as the limiting covariance matrix ${\bm\Sigma}$ is $\mathcal{F}_\infty$-measurable due to adaptive allocation. This stable convergence ensures joint convergence with respect to $\mathcal{F}_\infty$ \citep{Hall1980}, allowing random normings to be incorporated in the CLT.

    For statistical inference, let $\widehat{\bm{\Sigma}}_n$ be the plug-in estimator of $\bm{\Sigma}$ obtained by replacing $\sigma_k^2, C_{ks}, Z_k$ with the consistent estimators $\hat{\sigma}^2_{k,n}, \hat{C}_{ks,n}, \hat{Z}_{k,n}$ provided in Lemma \ref{lerr2} of Appendix \ref{appA}. Combined with the asymptotic vanishing of suboptimal allocation proportions established in Lemma \ref{lerr4}, and together with Lemmas \ref{ler5} and \ref{ler3}, this guarantees that
    \begin{equation*}
        \widehat{\bm{\Sigma}}_n^{-1/2} \begin{pmatrix}\sqrt{S_{n1}}(\hat\mu_{1,n}-\mu_1),\sqrt{S_{n2}}(\hat\mu_{2,n}-\mu_2) \dots, \sqrt{S_{nd}}(\hat\mu_{d,n}-\mu_d)\end{pmatrix}^\top
        \xrightarrow{\text{d}}{\cal N}({\bf 0}, {\bf I}_d),
    \end{equation*}
    where ${\bf I}_d$ is the $d \times d$ identity matrix.

	\subsection{Construction of the Test Statistics}\label{sec3.2}
    We next establish the asymptotic distribution for functions $h(\cdot)$ of the estimator of ${\bm\mu}$. To this end, we introduce notation for the arms involved in the hypothesis test. Let ${\mathcal{T}}_h\subseteq [1:d]$ denote the set of indices on which $h$ depends. Define $\mu_{h,\min}=\min\{\mu_k: k\in{\mathcal{T}}_h\}$ and ${\mathcal{T}}_{h,\min}=\{k:\mu_k=\mu_{h,\min},\ k\in{\mathcal{T}}_h\}$. For notational convenience, denote the partial derivative of $h$ with respect to $\mu_k$ by $\partial_k h(\cdot)$, and write $\partial_k h(\hat{\bm\mu}_n)$ for its evaluation at $\hat{\bm\mu}_n$.
    We assume $h$ satisfies the following assumptions.
    \begin{assumption}\label{asspar}
		The function $h: \mathbb{R}^d \to \mathbb{R}$ is continuously differentiable in a neighborhood of ${\bm\mu}$. Furthermore, there exists $k\in \mathcal{T}_{h,\min}$ such that $\partial_k h(\bm\mu) \neq 0$.
	\end{assumption}
	Theorem \ref{th3} establishes the asymptotic distribution of $h(\hat{\bm\mu}_n)$ under the UNB process.
	\begin{theorem}\label{th3}
		Suppose that Assumptions \ref{meanbound}, \ref{meanconver} and \ref{asspar} hold. Then, it holds that
		\begin{equation*}
			\frac{h(\hat{\bm\mu}_n)-h({\bm\mu})}{\hat{\sigma}_{h,n}} \xrightarrow{{\rm d}}{\mathcal{N}}(0,1),
		\end{equation*}
		where the variance estimator is
		$\hat{\sigma}^2_{h, n}=\sum_{i=1}^d \sum_{j=1}^d \frac{\partial_i h(\hat{\bm\mu}_n) \partial_j h(\hat{\bm\mu}_n)}{\sqrt{S_{ni}S_{nj}}} [\widehat{\bm{\Sigma}}_n]_{ij}$.
	\end{theorem}
	
	For example, the test statistic for hypothesis $H_0: h({\bm\mu}_{[{\mathcal{A}}]})\le0$ is
	\begin{equation}\label{eq:psi_n}
		\Psi_n = \frac{h(\hat{\bm\mu}_n)}{\hat{\sigma}_{h,n}}.
	\end{equation}
	Under the null hypothesis boundary, Theorem \ref{th3} implies that $\Psi_n \xrightarrow{{\rm d}} \mathcal{N}(0,1)$. Therefore, for a test with a significance level $\alpha$, the rejection region is given by ${\mathcal{C}}_\alpha=\{\Psi_n>z_{\alpha}\}$, where $z_\alpha$ is the upper $\alpha$-quantile of the standard normal distribution ${\mathcal{N}}(0,1)$.
	To understand the structural difference between our adaptive test and the classical two-sample $t$-test under fixed design, we examine the variance estimator $\hat{\sigma}^2_{h,n}$. It admits that 
	\begin{equation*}
        \hat{\sigma}^2_{h,n}=\hat{\Gamma}_n \cdot \frac{\sum_{k=1}^d {(\partial_k h(\hat{\bm\mu}_n))^2\hat{\sigma}^2_{k,n}}}{{S_{nk}}},\qquad\text{with}\quad
		\hat{\Gamma}_n=\frac{\sum_{i,j}\frac{\partial_i h(\hat{\bm\mu}_n)\partial_j h(\hat{\bm\mu}_n)}{\sqrt{S_{ni}S_{nj}}}[\widehat{\bm{\Sigma}}_n]_{ij}}{\sum_{k=1}^d \frac{(\partial_k h(\hat{\bm\mu}_n))^2\hat{\sigma}^2_{k,n}}{S_{nk}}}
	\end{equation*}
	The term $\sum_{k=1}^d {(\partial_k h(\hat{\bm\mu}_n))^2\hat{\sigma}^2_{k,n}}/{S_{nk}}$ corresponds to the variance component under the classical two-sample $t$-test, while $\hat{\Gamma}_n$ is the adaptive design correction factor. Lemma \ref{le4} shows the asymptotic behavior of $\hat{\Gamma}_n$.
	\begin{lemma}\label{le4}
		Suppose that Assumptions \ref{meanbound}, \ref{meanconver} and \ref{asspar} hold. If $\partial_k h(\bm\mu)\neq 0$ for some $k\notin \mathcal{I}$, then $\hat{\Gamma}_n\xrightarrow{\rm a.s.}1$. Otherwise, if $\partial_k h(\bm\mu)=0$ for all $k\notin \mathcal{I}$, then as $n\to\infty$,
		\begin{equation}\label{eqx1}
			\hat\Gamma_n\xrightarrow{{\rm a.s.}} \ \Gamma= 1+\frac{\sum_{i \neq j \in \mathcal{I}} \frac{\partial_i h({\bm\mu})\partial_j h({\bm\mu})}{\sqrt{Z_i Z_j}} C_{ij} f(N, Z_i, Z_j)}{\sum_{k \in \mathcal{I}} \frac{(\partial_k h({\bm\mu}))^2\sigma^2_k}{Z_k}}.
		\end{equation}
	\end{lemma}

    Lemma \ref{le4} characterizes the source of the variance correction in \eqref{eqx1}. When the function $h$ depends only on optimal arms, the adjustment term is entirely driven by their interaction. In this case, the allocation follows a multinomial sampling mechanism under a shared batch budget $N$, which induces dependence across arms. This structural effect is captured by the second term in \eqref{eqx1}, which depends on the gradient $\partial h(\bm\mu)$, the cross-arm covariances $C_{ij}$, the limiting proportions $Z_k$, and the overlap term $\mathbb{E}[\min(X,Y)]$ arising from the multinomial allocation. Notably, in the single-play setting ($N_t\equiv 1$), mutually exclusive arm selection ensures $\min(X,Y)=0$ almost surely, yielding $\Gamma=1$.

    \begin{figure}[htbp]
		\centering
		\includegraphics[width=0.9\textwidth]{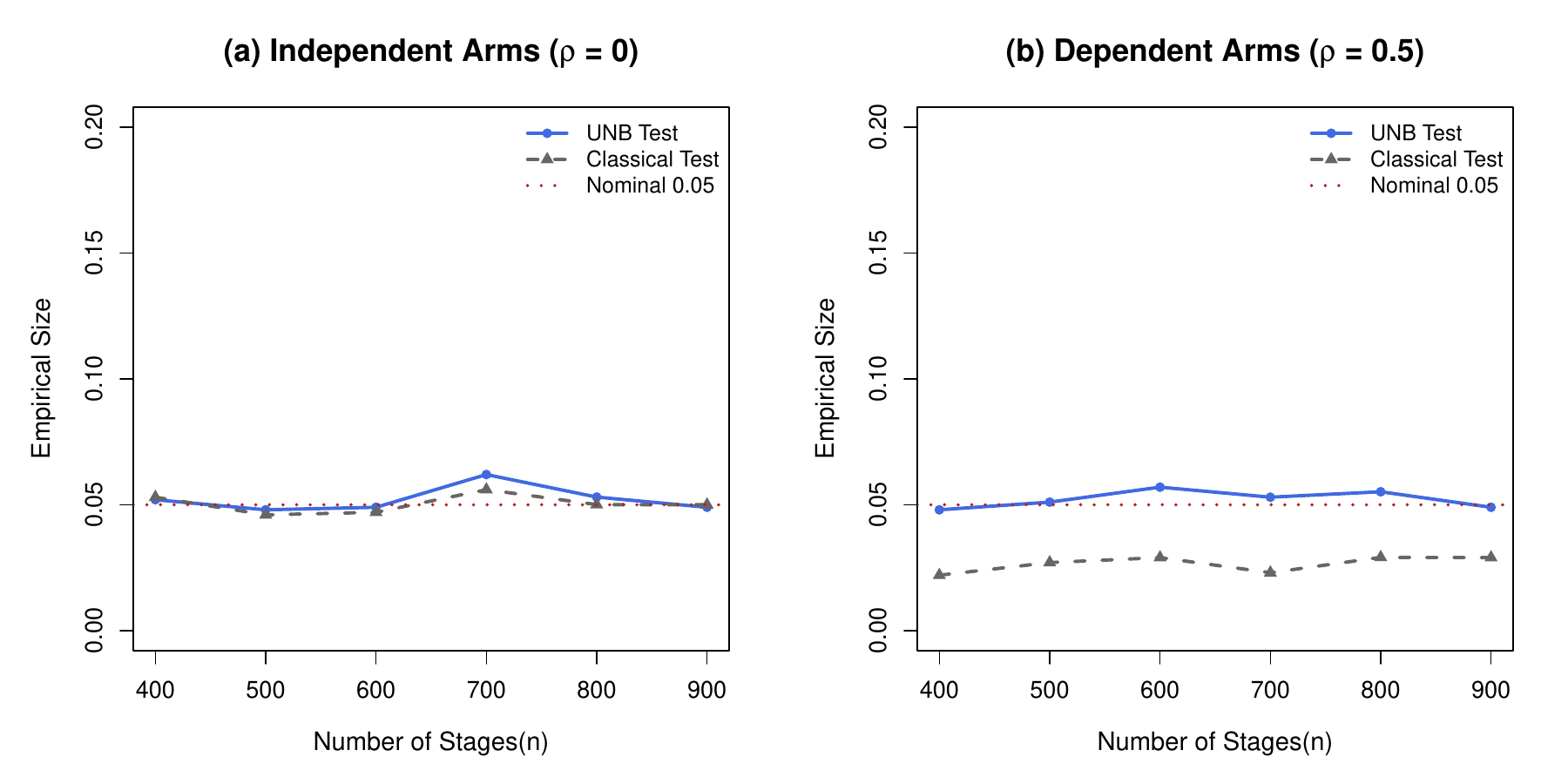}
		\vspace{-0.5cm}
		\caption{Empirical Size of the UNB test and the naive classical test under the null $H_0: \mu_1=\mu_2$ with batch size of $N_t=4$ across different cross-arm dependence. }
		\label{tu1}
	\end{figure}

    Figure~\ref{tu1} reports the empirical size under $H_0:\mu_1=\mu_2$ across different levels of cross-arm reward dependence, parameterized by the correlation coefficient $\rho$. When the cross-arm rewards are independent ($\rho=0$, Figure~\ref{tu1}(a)), both the classical test and $\Psi_n$ control size at the nominal level. This is consistent with the theoretical result that in the absence of cross-arm dependence, the variance estimator of the $t$-test remains asymptotically valid despite the adaptive data collection. When the rewards exhibit cross-arm dependence (e.g., $\rho=0.5$, Figure~\ref{tu1}(b)), the classical test exhibits size distortion, whereas $\Psi_n$ maintains the nominal level. This is due to the explicit incorporation of the cross-arm covariance structure, which becomes essential when cross-arm dependence interacts with adaptive sampling.

	\subsection{Asymptotic Power Analysis}
	This section explores the asymptotic behavior under fixed alternatives and characterizes the asymptotic power of the test. Notably, the divergence rate of the test statistic is governed by $\mu_{h,\min}/(2\mu^\ast)$ in Theorem \ref{thrr2}, indicating that the test's asymptotic efficiency is constrained by the arm with the minimum effective sample size.

	\begin{theorem}\label{thrr2}
		Suppose that Assumptions \ref{meanbound}, \ref{meanconver} and \ref{asspar} hold. Under any fixed alternative $H_1:h({\bm\mu}_{[{\mathcal{A}}]})=\tau>0$, conditional on the limiting scaled allocation proportions of the arms, the distribution of $\Psi_n$ is asymptotically normal with unit variance and a mean that diverges at a rate of $n^{\frac{\mu_{h,\min}}{2\mu^\ast}}$. Thus, 
		\begin{equation*}
			\mathbb{P}_{H_1}({\rm Reject}\ H_0)=\mathbb{P}_{H_1}(\Psi_n>z_\alpha)\to1,\quad \text{as}\ n\to\infty.
		\end{equation*}
	\end{theorem}

	To contextualize the asymptotic statistical efficiency of the UNB test, we compare it with two widely used allocation benchmarks: equal randomization (ER) and the UCB algorithm. ER is a non-adaptive benchmark that maximizes power, and the UCB represents a strategy optimized for regret minimization. 

    A principled comparison evaluates each strategy as a complete data-generating process. In adaptive settings, sample sizes are endogenously determined by the allocation rule, so inference efficiency is tied to the allocation mechanism. We therefore adopt three principles: (i) each strategy generates its own observation sequence; (ii) each strategy is evaluated using its valid test statistic to ensure Type I error control; and (iii) all operate under identical constraints, namely a fixed sample budget and a nominal level $\alpha=0.05$. 
    For inference, we employ $\Psi_n$ for UNB, the classical $Z$-statistic $\Psi_n^{\mathrm{ER}}$ for ER, and a $Z$-type statistic for UCB motivated by recent asymptotic normality results (e.g., \cite{Khamaru2024}). The explicit forms under the testing scenarios in Section \ref{sec2.2} are summarized in Table \ref{tab4}, where $S_{nk}, T_{nk}$ and $n_k$ denote the arm-specific sample sizes under UNB, UCB and ER, respectively. The correction factors $\hat\Gamma^{(j)}_n$ are given by
    \begin{align*}
        \hat\Gamma^{(1)}_n=1-\frac{ \frac{2[\widehat{\bm{\Sigma}}_n]_{12} }{\sqrt{S_{n1}S_{n2}}}}{ \frac{\hat{\sigma}^2_{1,n}}{S_{n1}}+\frac{\hat{\sigma}^2_{2,n}}{S_{n2}} }, \quad 
        \hat\Gamma^{(2)}_n=1,\quad \hat\Gamma^{(3)}_n = 1+\frac{ - \frac{[\widehat{\bm{\Sigma}}_n]_{12}}{\sqrt{S_{n1}S_{n2}}}-\frac{[\widehat{\bm{\Sigma}}_n]_{13}}{\sqrt{S_{n1}S_{n3}}}+\frac{0.5 [\widehat{\bm{\Sigma}}_n]_{23}}{\sqrt{S_{n2}S_{n3}}}}{ \frac{\hat{\sigma}^2_{1,n}}{S_{n1}} + \frac{0.25\hat{\sigma}^2_{2,n}}{S_{n2}} + \frac{0.25\hat{\sigma}^2_{3,n}}{S_{n3}}}.
    \end{align*}
    
	\begin{table}[htbp]
		\centering
        \renewcommand{\arraystretch}{1.8}
		\caption{Summary of test statistics applied by allocation strategies UNB, UCB, ER to hypothesis tests in Section \ref{sec2.2}.}
		\label{tab4}
		\scalebox{0.9}{
			\begin{tabular}{lccc}
				\toprule
				\textbf{Hypothesis Test} & \textbf{UNB ($\Psi_n$)} & \textbf{UCB ($\Psi_n^{\text{UCB}}$)} & \textbf{ER ($\Psi_n^{\text{ER}}$)} \\
				\midrule
				$H^{(1)}_0: \mu_1 \le \mu_2$ & $\frac{\hat{\mu}_{1,n} - \hat{\mu}_{2,n}}{\sqrt{\hat\Gamma^{(1)}_n}\sqrt{\frac{\hat{\sigma}^2_{1,n}}{S_{n1}} + \frac{\hat{\sigma}^2_{2,n}}{S_{n2}}}}$ & 
				$\frac{\hat{\mu}_{1,n} - \hat{\mu}_{2,n}}{\sqrt{\frac{\hat{\sigma}^2_{1,n}}{T_{n1}} + \frac{\hat{\sigma}^2_{2,n}}{T_{n2}}}}$ & 
				$\frac{\hat{\mu}_{1,n} - \hat{\mu}_{2,n}}{\sqrt{\frac{\hat\sigma^2_{1,n}}{n_k} + \frac{\hat\sigma^2_{2,n}}{n_k}}}$ \\[2ex] 
				$H^{(2)}_0: \mu_k \le K_0$ & $\frac{\hat{\mu}_{k,n} - K_0}{\sqrt{\hat\Gamma^{(2)}_n}\sqrt{\frac{\hat{\sigma}^2_{k,n}}{S_{nk}}}}$ &  $\frac{\hat{\mu}_{k,n}-K_0}{\sqrt{\frac{\hat{\sigma}^2_{k,n}}{T_{nk}}}}$ & $\frac{\hat{\mu}_{k,n}-K_0}{\sqrt{\frac{\hat\sigma^2_{k,n}}{n_k}}}$ \\[2ex]
				$H^{(3)}_0: \mu_1\le\frac{\mu_2+\mu_3}{2}$ & 
				$\frac{\hat{\mu}_{1,n}-0.5(\hat{\mu}_{2,n} + \hat{\mu}_{3,n})}{\sqrt{\hat\Gamma^{(3)}_n}\sqrt{\frac{\hat{\sigma}^2_{1,n}}{S_{n1}} + \frac{0.25\hat{\sigma}^2_{2,n}}{S_{n2}} + \frac{0.25\hat{\sigma}^2_{3,n}}{S_{n3}}}}$ &
				$\frac{\hat{\mu}_{1,n} - 0.5(\hat{\mu}_{2,n} + \hat{\mu}_{3,n})}{\sqrt{\frac{\hat{\sigma}^2_{1,n}}{T_{n1}} + \frac{0.25\hat{\sigma}^2_{2,n}}{T_{n2}} + \frac{0.25\hat{\sigma}^2_{3,n}}{T_{n3}}}}$ & $\frac{\hat{\mu}_{1,n}-0.5(\hat{\mu}_{2,n} + \hat{\mu}_{3,n})}{\sqrt{\frac{\hat\sigma^2_{1,n}}{n_k} + \frac{0.25\hat\sigma^2_{2,n}}{n_k}+\frac{0.25\hat\sigma^2_{3,n}}{n_k}}}$ \\
				\bottomrule
			\end{tabular}
		}
	\end{table}

	With the explicit forms of the test statistics established, we now proceed to a theoretical comparison of their asymptotic powers. While the proposed framework applies to general hypotheses, the fundamental statistical efficiency gap between the strategies is most clearly illuminated in a direct pairwise comparison.  Next, we focus our asymptotic analysis on the classical two-arm setup involving a suboptimal arm:
	\begin{equation}\label{eqtwoarm}
		H_0: \mu_1 \le \mu_2\quad {\rm versus}\quad  H_1: \mu_1-\mu_2=\Delta>0.
	\end{equation}
	
	Under the alternative $H_1$ in (\ref{eqtwoarm}), the test statistics ${\Psi}^{(\cdot)}_n$ are asymptotically normal with mean $\text{NCP}_n^{(\cdot)}$, where the non-centrality parameter (NCP) is induced by the allocation strategy. For the ER benchmark that maximizes power, the NCP scales at the optimal rate of $\text{NCP}^{\text{ER}}_n \asymp \sqrt{n}$. For the adaptive strategies, we have
	\begin{align}\label{eqncp}
		\text{NCP}^{\text{UNB}}_n=\frac{1}{\sqrt{\widehat{\Gamma}_n^{(1)}}}\frac{\Delta}{\sqrt{\frac{\widehat{\sigma}_{1, n}^2}{S_{n1}}+\frac{\widehat{\sigma}_{2,n}^2}{S_{n2}}}},\quad
		\text{NCP}^{\text{UCB}}_n=\frac{\Delta}{\sqrt{\frac{\widehat{\sigma}_{1,n}^2}{T_{n1}}+\frac{\widehat{\sigma}_{2,n}^2}{T_{n2}}}}.
	\end{align}
	By characterizing the asymptotic growth of the effective sample sizes under each allocation rule, we establish the divergence rates of $\text{NCP}^{(\cdot)}_n$.
	\begin{corollary}\label{cocom}
		Suppose that Assumptions \ref{meanbound}, \ref{meanconver} and \ref{asspar} hold. In addition, assume that the rewards are sub-Gaussian. Under the fixed alternative $H_1:\mu_1-\mu_2=\Delta>0$, the ${\rm NCP}^{(\cdot)}_n$ defined in \eqref{eqncp} satisfies that
        ${\rm{NCP}}^{\rm{UNB}}_n \asymp \sqrt{n^{\mu_2/\mu^\ast}}$ and ${\rm{NCP}}^{\rm{UCB}}_n \asymp \sqrt{\log n}$. Since the power is monotonically increasing in ${\rm NCP}_n^{(\cdot)}$, it follows that for all sufficiently large $n$, ${\rm Power}^{{\rm UNB}}>{\rm Power}^{{\rm UCB}}$.
	\end{corollary}

\begin{figure}[htbp]
		\centering
		\includegraphics[width=0.8\textwidth]{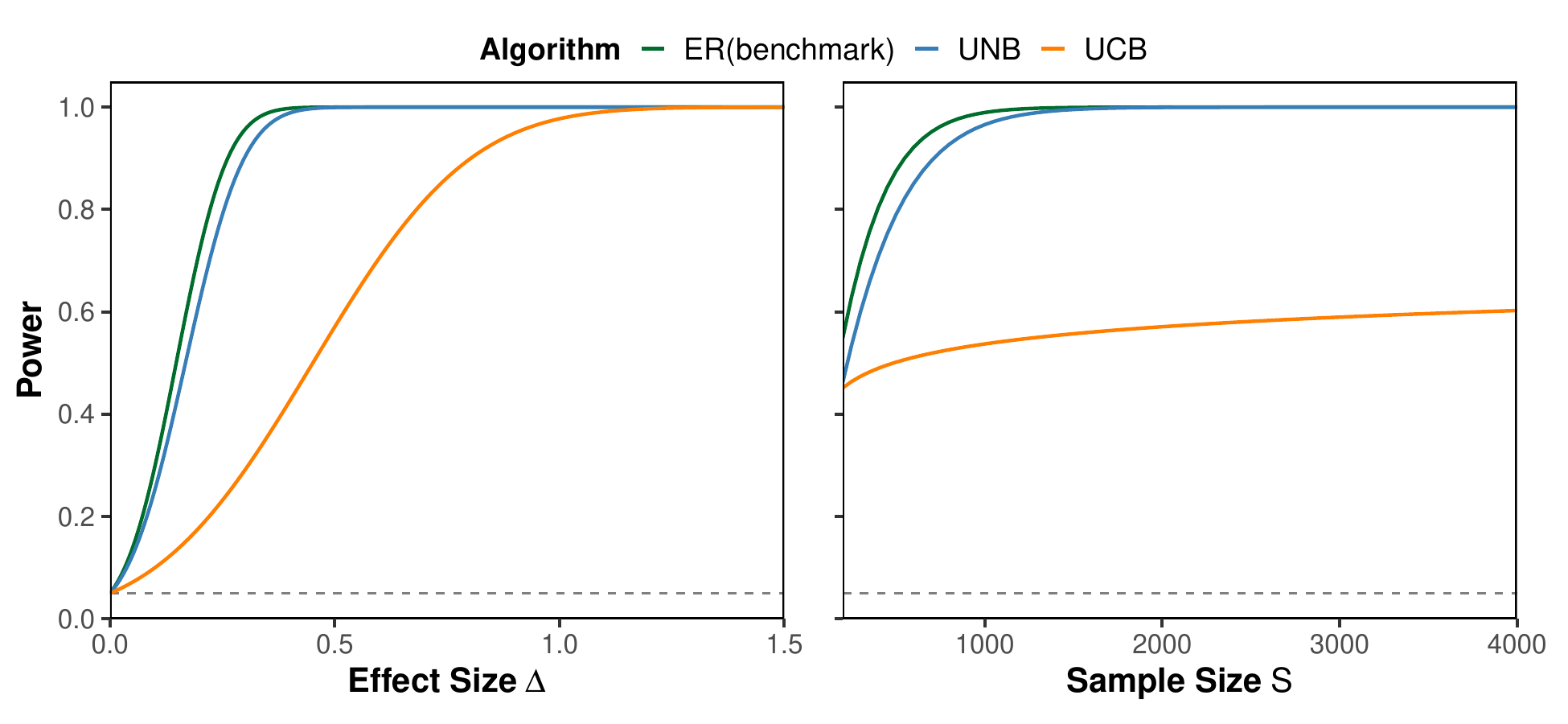}
		\vspace{-0.5cm}
		\caption{
			Asymptotic power curves of different allocation strategies for the two-arm test \eqref{eqtwoarm} across different $\Delta$ with $S=2000$ and different sample size $S$ at $\Delta=0.5$. UNB closely approximates the ER benchmark which maximizes power in both cases.
		}
		\label{power_curve}
	\end{figure}

    Corollary \ref{cocom} highlights a key difference in asymptotic power. 
    The NCP under UNB grows at a polynomial rate, while that under UCB grows only logarithmically. This reflects a trade-off in adaptive allocation, where stronger concentration improves exploitation but may limit information for the target test. As shown in Figure \ref{power_curve}, UNB's power closely tracks the ER benchmark. These theoretical results for asymptotic power extend directly to the other testing problems in Table \ref{tab4}.

	\vspace{-0.4cm}

    \section{Sequential tests under the UNB Process}\label{sec4}
	Embedding UNB within the sequential tests framework enables learning and early stopping to be carried out simultaneously, allowing the system to both exploit optimal arms and terminate once sufficient evidence against $H_0$ has accumulated. However,
	(i) Repeated interim analyses require explicit control of the dependence among interim statistics to maintain the nominal Type I error \citep{ArmitageMcPhersonRowe1969};
	(ii) Classical sequential tests typically rely on Gaussian approximations variance stabilized scales \citep{Jennison2000}. Yet, UNB exhibits stochastic and heterogeneous variance accumulation: optimal arms grow linearly while suboptimal arms scale sublinearly. This path dependent non-uniformity obscures the appropriate scaling and precludes direct calendar time boundaries.
	
	\subsection{Functional Central Limit Theorem for the UNB Process}\label{sec4.1}
	We extend fixed-time asymptotics to an FCLT for the vector of mean estimators generated by the UNB process. A key feature is the heterogeneous accumulation of effective sample size across arms, where $S_{nk}=O_{{\rm a.s.}}(n)$ for optimal arms and $S_{nk}=O_{{\rm a.s.}}(n^{\mu_k/\mu^\ast})$ for suboptimal arms. 
	\begin{theorem}[FCLT with Stable Convergence]\label{th1}
		Suppose that Assumptions \ref{meanbound}, \ref{meanconver} and \ref{asspar} hold. Define the cumulative deviation process scaled by the total sample size{\rm:}
		\begin{equation*}
			{\bf M}_n(t)=\left(\frac{S_{\lfloor nt \rfloor, 1}}{\sqrt{S_{n1}}}(\hat\mu_{1,\lfloor nt\rfloor}-\mu_1), \dots, \frac{S_{\lfloor nt \rfloor, d}}{\sqrt{S_{nd}}}(\hat\mu_{d,\lfloor nt\rfloor}-\mu_d)\right)^\top, \quad t \in [0,1].
		\end{equation*}
		As $n\to\infty$, the process converges stably in the Skorokhod space ${\mathcal{D}}([0,1],\mathbb{R}^d)${\rm:}
		\begin{equation*}
			{\bf M}_n(\cdot) \xrightarrow{{\rm d(stably)}} {\bf G}(\cdot),
		\end{equation*}
		where the conditional covariance structure of ${\bf G}(\cdot)$ is given by
		${\mathbb{C}\text{ov}}({\bf G}(t)\mid \mathcal{F}_\infty)={\bf D}(t)\,{\bm\Sigma}\,{\bf D}(t)$, with the time-scaling matrix ${\bf D}(t)={\rm diag}\big(t^{\frac{\mu_1}{2\mu^\ast}}, \dots, t^{\frac{\mu_d}{2\mu^\ast}}\big)$ and ${\bm\Sigma}$ as defined in Theorem \ref{th2}.
	\end{theorem}
	
	The diagonal matrix ${\bf D}(t)$ formalizes the impact of heterogeneous sampling rates on the fluctuation process. For optimal arms, the exponent $1/2$ recovers the canonical $\sqrt{t}$ scaling of standard Brownian motion, while for suboptimal arms, the smaller exponent $\mu_k/(2\mu^\ast) < 1/2$ compresses their effective time-scale, dampening variance accumulation relative to the optimal arms.

	\subsection{Information Fraction Transformation and Canonical Joint Distribution}\label{sec4.2}

    To ensure valid sequential inference under sublinear growth  $n^{\mu_k/\mu^\ast}$, we adopt the information fraction framework \citep{Jennison2000}. The sequential test statistic is indexed by accumulated observed information $I_n=(\widehat\sigma_{h,n}^2)^{-1}$, which denotes the inverse variance of the adaptive estimator. This formulation with $I_0=0$ scales the analysis by estimation precision, where a larger $I_n$ reflects reduced variance and a more precise estimate of $h({\bm\mu})$ \citep{Proschan2024,LanZucker1993}.

	We index the test process on $[0,1]$ by the calendar time fraction $r$ and define the information fraction relative to the terminal level $
		t_n(r)=\frac{I_{\lfloor nr\rfloor}}{I_n}$.
	This yields $t_n(0)=0$ and $t_n(1)=1$. Recall $\mu_{h,\min}=\min\{\mu_k:k\in{\mathcal{T}}_h\}$ and define $\gamma=\frac{\mu_{h,\min}}{\mu^\ast}$.
	The following lemma characterizes the asymptotic scaling of the information fraction relative to the calendar time fraction for the testing problem \eqref{eq3} under UNB.
	\begin{lemma}\label{le6}
		Suppose that Assumptions \ref{meanbound}, \ref{meanconver} and \ref{asspar} hold. For any calendar time $0\le r< s\le1$,
		\begin{equation}\label{eq12}
			\frac{t_n(r)}{t_n(s)}\xrightarrow{{\rm a.s.}}\left(\frac{r}{s}\right)^\gamma,\qquad \text{as } n\to\infty.
		\end{equation}
	\end{lemma}
	
	Lemma \ref{le6} implies a regular-variation type relationship between information fraction and the calendar time fraction. Applying \eqref{eq12} with $s=1$ and $t_n(1)\equiv1$, we obtain
	$ t_n(r)\xrightarrow{{\rm a.s.}} r^\gamma$
	for each fixed $r\in[0,1]$. Denoting this limit function by $t(r)=r^\gamma$, it is strictly increasing on $[0,1]$ and admits the inverse mapping $g(t)=t^{\frac{1}{\gamma}}$. This motivates a reparameterization of the test process to recover the canonical Brownian covariance structure. Utilizing the inverse mapping to identify the calendar time corresponding to the information fraction $t$, we define the information fraction indexed process $B_n(t)$ for $t\in [0,1]$ as $B_n(0)=0$ and
	\begin{equation*}
		B_n(t)=\sqrt{t} \frac{h(\hat{\bm\mu}_{\lfloor ng(t)\rfloor})-h(\bm\mu)}{\hat{\sigma}_{h,{\lfloor ng(t)\rfloor}}}, \quad \text{for\ \ }  t \in (0,1].
	\end{equation*}
	The reparametrization above restores the standard Brownian motion limit.
	
	\begin{theorem}[FCLT under information fraction]\label{th6}
		Suppose that Assumptions \ref{meanbound}, \ref{meanconver} and \ref{asspar} hold. As $n\to\infty$, the process $B_n(\cdot)$ satisfies that
		\begin{equation*}
			B_n(\cdot)\xrightarrow{\text{d}}\mathbb W(\cdot),
		\end{equation*}
		in ${\mathcal{D}}([0,1])$, where $\mathbb W(\cdot)$ is a standard Brownian motion. 
	\end{theorem}

    Since data are observed in calendar time $r$, we define the sequential statistic as
	\begin{equation*}\label{eqphi}            
    \Psi_n(r)=\frac{h(\hat{\bm\mu}_{\lfloor nr\rfloor})}{\hat\sigma_{h,\lfloor nr\rfloor}}.
	\end{equation*}
	
	The Brownian motion limit in Theorem \ref{th6} yields the joint distribution of the standardized statistics at any finite collection of information fractions. In particular, at the boundary of the null hypothesis $h(\bm\mu_{[{\mathcal{A}}]})=0$, evaluating $B_n(\cdot)$ at $t_1,\dots,t_K$ with $\Psi_n(g(t))=\frac{B_n(t)}{\sqrt{t}}$ yields the following corollary.
	\begin{corollary}\label{co2}
		Suppose that Assumptions \ref{meanbound}, \ref{meanconver} and \ref{asspar} hold. Let $0 < t_1 < \dots < t_K \le 1$ be a fixed sequence of information fractions. Under the null hypothesis boundary $h(\bm\mu_{[{\mathcal{A}}]})=0$, as $n \to \infty$, the sequential test statistic
		\begin{equation*}
			\left(\Psi_n(g(t_1)),\Psi_n(g(t_2)),\cdots,\Psi_n(g(t_K)) \right)^\top
		\end{equation*}
		converges in distribution to a multivariate standard normal vector $(Z_1, \dots, Z_K)^\top$ with mean zero and the covariance structure $\mathbb{C}{\rm ov}(Z_i, Z_j)=\sqrt{\frac{t_{\min\{i,j\}}}{t_{\max\{i,j\}}}}$.
	\end{corollary}
	
	The sequential test statistic $\left(\Psi_n(g(t_1)),\Psi_n(g(t_2)),\cdots,\Psi_n(g(t_K)) \right)$ has the asymptotically {\it canonical joint distribution} defined in \cite[Chapter 3]{Jennison2000}. This result embeds the UNB process in the standard group sequential framework and provides the theoretical basis for using the $\alpha$ spending approach \citep{Lan1983} to construct stopping boundaries.

	Let $(c_1,\dots,c_K)$ denote a sequence of one-sided rejection boundaries. The following corollary shows the asymptotic power of the sequential test.
	\begin{corollary}\label{power_seq}
		Suppose that Assumptions \ref{meanbound}, \ref{meanconver} and \ref{asspar} hold. Under the fixed alternative $H_1: h(\bm\mu_{[{\mathcal{A}}]})=\tau>0$, the power of the sequential test converges to 1 as $n\to\infty$. Formally,
		\begin{equation*}
			\mathbb{P}_{H_1}({\rm Reject}\ H_0)=\mathbb{P}_{H_1}\big(\exists 
			\ k\le K: \Psi_n(g(t_k))>c_k \big)\to 1.
		\end{equation*}
	\end{corollary}

	\subsection{Information Planning and Boundary Construction}\label{sec4.3}
	
	Sequential tests require two design components: (i) Information planning, which specifies the target information level ${I}_{\max}$ to attain the power $1-\eta$ at significance level $\alpha$; (ii) Stopping boundary construction, which derives critical values via an $\alpha$ spending approach to control the overall Type I error.
	Following \cite{LanZucker1993}, we adopt a normal model for the test statistic in \eqref{eq:psi_n} and calibrate ${ I}_{\max}$ accordingly, yielding
	${{I}}_{\max}=\left(\frac{z_\alpha+z_\eta}{\Delta}\right)^2$.
	To account for repeated interim analyses, we inflate the target information to $\tilde{ I}_{\max}=L{ I}_{\max}$, where $L>1$ is set by the number of looks $K$ and the $\alpha$ spending approach \citep{Jennison2000}. Interim analyses occur as observed information first exceeds pre-specified fractions of $\tilde{I}_{\max}$.

	To control the overall Type I error at level $\alpha$, we fix a non-decreasing $\alpha$ spending function $\alpha^\ast:[0,1]\to[0,\alpha]$ with $$\alpha^\ast(0)=0,\quad \alpha^\ast(1)=\alpha.$$ At the $k$th analysis with information fraction $t_k$, the incremental error is $\Delta\alpha_k = \alpha^\ast(t_k)-\alpha^\ast(t_{k-1})$. The rejection boundaries $\{c_k\}$ are chosen so that, under $H_0$ and with respect to the limiting Gaussian vector $(Z_1,\dots,Z_K)^\top$ in Corollary \ref{co2},
	\begin{align*}
		\mathbb{P}(Z_1\ge c_1)=\Delta\alpha_1,\quad
		\mathbb{P}\big(Z_k\ge c_k,\ Z_j<c_j\ \text{for all }j<k\big)=\Delta\alpha_k,\qquad k=2,\dots,K.
	\end{align*}
	The stopping rule is $$\tilde\tau=\inf\{k\ge1:\Psi_{n_k}\ge c_k\},$$ and $H_0$ is rejected when $\tilde\tau\le K$. Algorithm \ref{alg:unb_seq} summarizes the sequential tests procedure.

    	The spending function $\alpha^\ast(\cdot)$ determines the shape of efficacy boundaries over information time. Choices include the Pocock--like, O'Brien--Fleming--like (OBF), power--type, and Hwang--Shih--DeCani (HS) families \citep{Lan1983}:
	\begin{align*}
		\alpha^\ast_1(t)&=\alpha\ln\big(1+(e-1)t\big),\quad  \alpha^\ast_2(t)=1-\Phi\left(\frac{z_\alpha}{\sqrt{t}}\right),\\
		\alpha^\ast_3(t)&=\alpha t^q,\quad \alpha^\ast_4(t)= \alpha \frac{1-e^{-\gamma t}}{1-e^{-\gamma}}.
	\end{align*}
	In Sections \ref{sec5} and \ref{sec6}, we adopt an OBF--like spending function, which yields conservative early boundaries relative to the accumulated information fraction and ensures rigorous control of the overall Type I error under adaptive allocation. 

       \begin{algorithm}[htbp]
		\caption{UNB allocation with information fraction group sequential tests}
		\label{alg:unb_seq}
		\begin{algorithmic}[1]
			
		\Require Significance level $\alpha$, target power $1-\eta$, number of looks $K$, information fraction $\tau^\ast_j=j/K$, $\alpha$ spending function, budget $\{N_t\}$, burn in $t_{\min}$.
	
		\State {\it Phase I: Design calibration}
		\State Compute $\tilde{I}_{\max}$ and critical values $\{c_j\}_{j=1}^K$ under the $\alpha$ spending function and information fraction $\{\tau^\ast_j\}_{j=1}^K$.

		\State {\it Phase II: Initialization}
		\State Set $j=1$ and initialize the cumulative statistics.

		\State {\it Phase III: Adaptive allocation and sequential test}
		
		\For{$t=1,2,\dots$}
		
		\State Draw ${\bf X}_t\sim{\rm Multinomial}(N_t;{\bf Z}_{t-1})$. Observe rewards $\{{\xi}_{tk,q}\}$ for drawn arms. 

		\State {\it Update cumulative sums and paired counts}
		\For{each arm $k$ and pair $k\ne s$}
		    \State $S_{tk}= S_{t-1,k}+X_{tk}$, $R_{tk}= R_{t-1,k}+\sum_{q=1}^{X_{tk}}\xi_{tk,q}$, $W_{tk}= W_{t-1,k}+\sum_{q=1}^{X_{tk}}\xi^2_{tk,q}$,
            \State
            $A_{ks,t}= A_{ks,t-1}+X_{tk}\wedge X_{ts}$, $B_{ks,t}= B_{ks,t-1}+\sum_{q=1}^{X_{tk}\wedge X_{ts}}\xi_{tk,q}\xi_{ts,q}$.   
		\EndFor

		\State {\it Update estimators in Lemma \ref{lerr2}}
		\State Compute ${\bf Z}_t= \frac{{\bf R}_t}{\|{\bf R}_t\|_1}$.
        \For{each arm $k$ and pair $k\ne s$}
        \State 
        $\hat{Z}_{k,t}=\frac{1}{t}[(t-1)\hat{Z}_{k,t-1}+X_{tk}/N_t]$, $\hat{\mu}_{k,t}= \frac{R_{tk}}{S_{tk}}$, $\hat{q}_{k,t}= \frac{W_{tk}}{S_{tk}}$, $\hat{\sigma}^2_{k,t}=\hat{q}_{k,t}-\hat{\mu}_{k,t}^2$,
        \State 
		   $\hat{q}_{ks,t}= \frac{B_{ks,t}}{A_{ks,t}}$, $\hat{C}_{ks,t}= \hat{q}_{ks,t}-\hat{\mu}_{k,t}\hat{\mu}_{s,t}$.
        \EndFor

		\State {\it Plug-in variance estimation as in Theorem~\ref{th3}}
		\State Construct $\widehat{\bm{\Sigma}}_t$ by Theorem~\ref{th2} and Lemma \ref{lerr2}. Set ${\bf b}_t= (\frac{\partial_1(h(\hat{\bm\mu}_t))}{\sqrt{S_{t1}}},\dots,\frac{\partial_d(h(\hat{\bm\mu}_t))}{\sqrt{S_{td}}})^\top$, $\hat{\sigma}^2_{h,t}={\bf b}_t^\top\widehat{\bm{\Sigma}}_t{\bf b}_t$, $\Psi_t=\frac{h(\hat{\bm\mu}_t)}{\hat{\sigma}_{h,t}}$. Set $I_t=\frac{1}{\hat{\sigma}^2_{h,t}}$ and $\tau_t= \frac{I_t}{\tilde{I}_{\max}}$.

		\State {\it Boundary checks at interim looks}
		\While{$t\ge t_{\min}$ \textbf{and} $j\le K$ \textbf{and} $\tau_t\ge \tau^\ast_j$}
		\If{$\Psi_t>c_j$}
		\State \Return Reject $H_0$ and stop.
		\EndIf
		\State $j= j+1$
		\EndWhile
			
		\If{$I_t\ge \tilde{I}_{\max}$}
		\State \Return Fail to reject $H_0$ and stop.
		\EndIf
			
		\EndFor
			
	\end{algorithmic}
    \end{algorithm}

	\section{Simulation Studies}\label{sec5}
	This section presents Monte Carlo studies of the finite sample performance of the proposed UNB, based on evaluation metrics including Type I error control, power, and early stopping behavior, compared with ER as a non-adaptive baseline and the UCB algorithm as a representative bandit baseline.

	\subsection{Fixed Sample Tests}\label{sec5.1}
	We first examine the fixed sample tests performance in the two-arm comparison across Bernoulli, Poisson, and Exponential reward distributions.
	We evaluate empirical size under the binding null $H_0:\mu_1=\mu_2$ and empirical power at a representative alternative with $\mu_1-\mu_2=\Delta$ ($\Delta>0$). In addition, we report the average number of observations assigned to the inferior arm, $S_{\text{inf}}$, to quantify how each allocation rule trades off precision for reduced exposure to the inferior arm.

    Under independent cross-arm rewards ($\rho=0$), Table \ref{sim_results} (top panel) shows that UNB maintains empirical size and achieves power comparable to the ER benchmark across all reward distributions. By allocating more observations to the optimal arm, UNB substantially reduces average $S_{\text{inf}}$. For instance, under Bernoulli $(0.6, 0.4)$, $S_{\text{inf}}$ drops from 135 (ER) to 89. This pattern persists for Poisson and Exponential outcomes, indicating significantly reduced inferior arm exposure without material power loss.
    For Bernoulli rewards, UCB attains near-nominal size and power comparable to UNB and ER. In contrast, it shows substantial size inflation for Poisson and Exponential outcomes.
    This may be attributed to the fact that the UCB index is derived from Hoeffding-type concentration inequalities, which are most accurate under sub-Gaussian assumptions. For Poisson and Exponential rewards, this calibration may be less appropriate, potentially leading to overly aggressive exploration and distorted allocation paths, which in turn affect the validity of the normal approximation underlying the test statistic.
    
    When cross-arm rewards are correlated ($\rho=0.5$), UNB maintains valid Type I error control while enabling adaptive allocation and early stopping, whereas ER and UCB exhibit inflated Type I error in the multi-draw setting, reflecting that their test statistics shown in Table~\ref{tab4} do not fully account for the covariance structure induced by cross-arm rewards and adaptive allocation.

	\subsection{Sequential Tests}\label{sec5.2}
	We next turn to the group sequential setting by employing the test procedure detailed in Algorithm \ref{alg:unb_seq} to allow for early stopping. Throughout, empirical size is evaluated under the null $H_0:\mu_1=\mu_2$, and simulations evaluating statistical efficiency and ethical performance are conducted under alternatives $H_1: \mu_1-\mu_2=\Delta$ ($\Delta>0$) calibrated to achieve a target power of $1-\eta=0.9$. 
    For each type of reward distribution, we report empirical size, power, and average sample number (ASN) used to attain the given power, alongside average inferior arm exposure $S_{\text{inf}}$ to quantify the ethical and statistical efficiency trade-off induced jointly by allocation and testing. 
	
	\begin{table}[H]
		\centering
		\caption{Simulated results for fixed sample and sequential tests under Bernoulli, Poisson, and Exponential reward distributions with arm independent $(\rho=0,N_n=1)$ and correlated $(\rho=0.5,N_n=4)$ settings. Empirical size is evaluated under $H_0$; power, $S_{\text{inf}}$, and ASN (for sequential tests only) are reported under $H_1$, with sequential tests calibrated to a target power of 0.9. ER is the power benchmark in the fixed sample test. Notation ``--'' implies that results are omitted  due to extreme Type I error inflation.}
		\label{sim_results}
		\scalebox{0.9}{
			\resizebox{1.06\linewidth}{!}{
		\setlength{\tabcolsep}{5pt} 
			\begin{tabular}{ll ccc ccc ccc ccc}
				\toprule
				& \multicolumn{13}{c}{\textbf{Bernoulli}} \\
				\cmidrule(lr){3-14}    & 
				$(\mu_1,\mu_2,\rho)\ \ \ S$    & \multicolumn{3}{c}{$(0.6,0.4,0)$\ \ 270} & \multicolumn{3}{c}{$(0.8,0.6,0)$\ \ 220} &  \multicolumn{3}{c}{$(0.6,0.4,0.5)$\ \ 200}  & \multicolumn{3}{c}{$(0.8,0.6,0.5)$\ \ 160} \\
				\cmidrule(lr){3-5} \cmidrule(lr){6-8} \cmidrule(lr){9-11}  \cmidrule(lr){12-14} & 
				& UNB & ER & UCB & UNB & ER & UCB &  UNB & ER & UCB & UNB & ER & UCB  \\
				\cmidrule{2-14}
				\multirow{18}{2cm}{Fixed Sample Tests}   & Emp. Size  & 0.049 & 0.052 & 0.052 & 0.049 & 0.051 & 0.048 & 0.050 & 0.015 & 0.022 & 0.049 & 0.021 & 0.017 \\
				& Emp. Power & 0.916 & 0.920 & 0.882 & 0.900 & 0.898 & 0.877 & 0.929 & {\bf --} & {\bf --} & 0.919 & {\bf --} & {\bf --} \\
				&  Average $S_{\rm inf}$ & 89 & 135 & 61 & 85 & 110 & 52 & 70 & {\bf --} & {\bf --} & 64 & {\bf --} & {\bf --} \\
				\cmidrule{2-14}
				& \multicolumn{13}{c}{\textbf{Poisson}} \\
				\cmidrule(lr){3-14}    & 
				$(\mu_1,\mu_2,\rho)\ \ \ S$   & \multicolumn{3}{c}{$(6.5,6,0)$\ \ 1040} & \multicolumn{3}{c}{$(11,10,0)$\ \ 460} &  \multicolumn{3}{c}{$(6.5,6,0.5)$\ \ 800} & \multicolumn{3}{c}{$(11,10,0.5)$\ \ 320}  \\
				\cmidrule(lr){3-5} \cmidrule(lr){6-8} \cmidrule(lr){9-11} \cmidrule(lr){12-14} & 
				& UNB & ER & UCB & UNB & ER & UCB & UNB & ER & UCB & UNB & ER & UCB \\
				\cmidrule{2-14}
				&   Emp. Size  & 0.052 & 0.048 & 0.119 & 0.049 & 0.049 & 0.099 & 0.052 & 0.02 & 0.03 & 0.051 & 0.023 & 0.029 \\
				&  Emp. Power & 0.914 & 0.904 & {\bf --} & 0.912 & 0.911 & {\bf --} & 0.919 & {\bf --} & {\bf --} & 0.917 & {\bf --} & {\bf --} \\
				&  Average $S_{\rm inf}$ & 457 & 520 & {\bf --} & 202 & 230 & {\bf --} & 254 & {\bf --} & {\bf --} & 109 & {\bf --} & {\bf --} \\
				\cmidrule{2-14}
				& \multicolumn{13}{c}{\textbf{Exponential}} \\
				\cmidrule(lr){3-14} &
				$(\mu_1,\mu_2,\rho)\ \ \ S$   & \multicolumn{3}{c}{$(7.5,6,0)$\ \ 900} & \multicolumn{3}{c}{$(12,10,0)$\ \ 1400}  & \multicolumn{3}{c}{$(7.5,6,0.5)$\ \ 640} & \multicolumn{3}{c}{$(12,10,0.5)$\ \ 960} \\
				\cmidrule(lr){3-5} \cmidrule(lr){6-8} \cmidrule(lr){9-11} \cmidrule(lr){12-14} &
				& UNB & ER & UCB & UNB & ER & UCB & UNB & ER & UCB & UNB & ER & UCB \\
				\cmidrule{2-14}
				&   Emp. Size  & 0.050 & 0.049 & 0.203 & 0.048 & 0.049 & 0.202 & 0.050 & 0.027 & 0.036 & 0.052 & 0.020 & 0.024 \\
				&   Emp. Power & 0.900 & 0.914 & {\bf --} & 0.902 & 0.911 & {\bf --} & 0.906 & {\bf --} & {\bf --} & 0.913 & {\bf --} & {\bf --} \\
				&   Average $S_{\rm inf}$ & 308 & 450 & {\bf --} & 496 & 700 & {\bf --} & 231 & {\bf --} & {\bf --} & 354 & {\bf --} & {\bf --} \\
				\cmidrule(lr){1-14} 
				& \multicolumn{13}{c}{\textbf{Bernoulli}} \\
				\cmidrule(lr){3-14}    & 
				$(\mu_1,\mu_2,\rho)$    & \multicolumn{3}{c}{$(0.6,0.4,0)$} & \multicolumn{3}{c}{$(0.8,0.6,0)$} &  \multicolumn{3}{c}{$(0.6,0.4,0.5)$}  & \multicolumn{3}{c}{$(0.8,0.6,0.5)$} \\
				\cmidrule(lr){3-5} \cmidrule(lr){6-8} \cmidrule(lr){9-11}  \cmidrule(lr){12-14} & 
				& UNB & ER & UCB & UNB & ER & UCB &  UNB & ER & UCB & UNB & ER & UCB  \\
				\cmidrule{2-14}
				\multirow{21}{2cm}{Sequential Tests}   & Emp. Size  & 0.052 & 0.051 & 0.053 & 0.049 & 0.048 & 0.048  & 0.049 & 0.007 & 0.037 & 0.050 & 0.010 & 0.030 \\
				& Emp. Power & 0.914 & 0.900 & 0.905 & 0.921 & 0.901 & 0.905  & 0.921  & {\bf --} & {\bf --} & 0.884 & {\bf --} & {\bf --} \\
				&   ASN   & 149 & 140 & 181 & 125 & 115 & 174  & 110 & {\bf --} & {\bf --} & 96 & {\bf --} & {\bf --} \\
				&  Average $S_{\rm inf}$ & 58 & 70 & 47 & 54 & 57 & 44 & 45 & {\bf --} & {\bf --} & 42 & {\bf --} & {\bf --} \\
				\cmidrule{2-14}
				& \multicolumn{13}{c}{\textbf{Poisson}} \\
				\cmidrule(lr){3-14}    & 
				$(\mu_1,\mu_2,\rho)$   & \multicolumn{3}{c}{$(6.5,6,0)$} & \multicolumn{3}{c}{$(11,10,0)$} &  \multicolumn{3}{c}{$(6.5,6,0.5)$} & \multicolumn{3}{c}{$(11,10,0.5)$}  \\
				\cmidrule(lr){3-5} \cmidrule(lr){6-8} \cmidrule(lr){9-11} \cmidrule(lr){12-14} & 
				& UNB & ER & UCB & UNB & ER & UCB & UNB & ER & UCB & UNB & ER & UCB \\
				\cmidrule{2-14}
				&   Emp. Size  & 0.049 & 0.048 & 0.004 & 0.050 & 0.049 & 0.007 &  0.048  & 0.007 & 0.023 & 0.051 & 0.011 & 0.032 \\
				&  Emp. Power & 0.901 & 0.889 & {\bf --} & 0.908 & 0.892 & {\bf --}  & 0.908 & {\bf --} & {\bf --} & 0.906 & {\bf --} & {\bf --} \\
				&  ASN        & 610 & 610 & {\bf --} & 251 & 248 & {\bf --}  & 406 & {\bf --} & {\bf --} & 174 & {\bf --} & {\bf --} \\
				&  Average $S_{\rm inf}$ & 276 & 305 & {\bf --} & 116 & 124 & {\bf --}  & 186 & {\bf --} & {\bf --} & 81 & {\bf --} & {\bf --} \\
				\cmidrule{2-14}
				& \multicolumn{13}{c}{\textbf{Exponential}} \\
				\cmidrule(lr){3-14} &
				$(\mu_1,\mu_2,\rho)$   & \multicolumn{3}{c}{$(7.5,6,0)$} & \multicolumn{3}{c}{$(12,10,0)$}  & \multicolumn{3}{c}{$(7.5,6,0.5)$} & \multicolumn{3}{c}{$(12,10,0.5)$} \\
				\cmidrule(lr){3-5} \cmidrule(lr){6-8} \cmidrule(lr){9-11} \cmidrule(lr){12-14} &
				& UNB & ER & UCB & UNB & ER & UCB & UNB & ER & UCB & UNB & ER & UCB \\
				\cmidrule{2-14}
				&    Emp. Size  & 0.052 & 0.050 & 0.026 & 0.048 & 0.049 & 0.020  & 0.048 & 0.005 & 0.062 & 0.050 & 0.007 & 0.039 \\
				&   Emp. Power & 0.903 & 0.900 & {\bf --} & 0.899 & 0.890 & {\bf --}  & 0.900 & {\bf --} & {\bf --}  & 0.905 & {\bf --} & {\bf --} \\
				&  ASN        & 491 & 483 & {\bf --} & 748 & 731 & {\bf --}  & 369 & {\bf --} & {\bf --} & 549 & {\bf --} & {\bf --}\\
				&    Average $S_{\rm inf}$ & 191 & 242 & {\bf --} & 295 & 365 & {\bf --}  & 145 & {\bf --} & {\bf --} & 215 & {\bf --} & {\bf --} \\
				\bottomrule
			\end{tabular}
		}}
	\end{table}

	Table \ref{sim_results} (bottom panel) summarizes these results. 
	Under the cross-arm independence ($\rho=0$), the 
	simulated results imply that the UNB process maintains empirical size close to the nominal level and power similar to the power benchmark ER across all scenarios under the target power $0.9$. Meanwhile, UNB consistently reduces inferior arm exposure $S_{\rm inf}$, yielding more rewards than ER. Specifically, for Exponential rewards, UNB reduces the average $S_{\rm inf}$ from $242$ (ER) to $191$ at $(7.5,6)$ and from $365$ to $295$ at $(12,10)$. Similar reductions occur for Bernoulli and Poisson rewards.
    The corresponding ASNs under UNB are close to ER with smaller average $S_{\rm inf}$, suggesting that the exposure reductions are not necessarily obtained by materially increasing expected sample size. 
    The Bernoulli results for UCB additionally highlight the statistical efficiency cost of aggressive adaptation. While UCB attains comparable size and power, it yields a markedly larger ASN than UNB or ER, e.g., $174$ vs. $125$ or $115$ for the Bernoulli reward at $(0.8,0.6)$. On the other hand, 
	the empirical size of UCB is not properly controlled under Poisson and Exponential rewards and is omitted due to extreme Type I error inflation. 
	
	Table \ref{sim_results} (bottom panel) also demonstrates that the UNB based sequential tests remain valid under cross-arm dependence where multiple arms are sampled simultaneously. Across all scenarios with $\rho=0.5$, the empirical size is well controlled around the nominal level. Under $H_1$, the empirical power remains close to the target level $0.9$, indicating that dependence does not distort the effective information accumulation required for sequential stopping. Moreover, a positive correlation $\rho>0$ can reduce the variance of the estimated mean difference under paired sampling, thereby accelerating information accumulation, leading to a smaller required ASN and a corresponding reduction in inferior arm exposure $S_{\rm inf}$ relative to the independent case.
    Since ER and UCB exhibit extreme Type I error inflation under fixed sample tests with cross-arm dependence, they are excluded from the sequential analysis in Table \ref{sim_results}, which aims to support valid sequential inference under adaptive allocation. 
	
	\begin{figure}[htbp]
		\centering
		\includegraphics[width=0.9\textwidth]{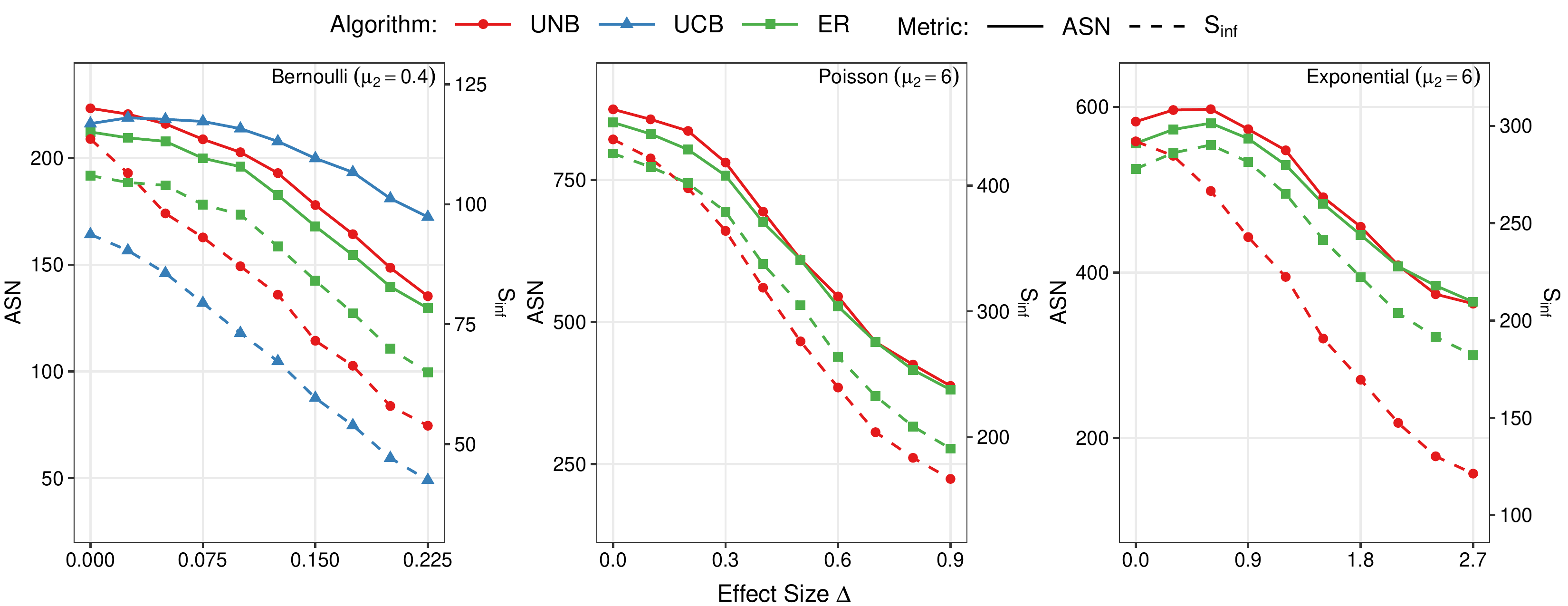}
		\vspace{-0.5cm}
		\caption{Dual-axis plots of ASN (left axis, solid lines) and $S_{\text{inf}}$ (right axis, dashed lines) versus $\Delta$ under the information-based sequential design. The baseline means are set to $0.4$, $6$, and $6$ for the Bernoulli, Poisson, and Exponential rewards, respectively. UNB balances ethics and statistical efficiency in terms of the similar ASN to ER but smaller $S_{\text{inf}}$. The results for UCB under Poisson and Exponential rewards are omitted due to extreme Type I error inflation.} 
		\label{tu2}
	\end{figure}
	
    Figure \ref{tu2} reports ASN and $S_{\text{inf}}$ across different $\Delta$, illustrating how early stopping and inferior arm exposure vary with reward effect size. Under an information-based sequential design, for a given $\Delta$, power is primarily driven by accumulated information and is therefore approximately invariant across algorithms that ensure valid inference on the same information scale. We thus focus on ASN and $S_{\text{inf}}$ as primary metrics.
    ASN decreases with $\Delta$, reflecting more frequent early stopping under stronger signals, with mild nonmonotonicity at small gaps due to discrete interim looks and early allocation imbalance. Meanwhile, $S_{\rm inf}$ also decreases with $\Delta$, indicating that larger effects reduce both stopping time and inferior arm allocation.
    Across all settings, UNB yields ASN trajectories comparable to those of ER while achieving a smaller $S_{\rm inf}$, thereby improving ethical performance at only a minor ASN cost. In the Bernoulli case, UCB minimizes inferior arm exposure at the cost of a higher ASN, as extreme allocation imbalance slows information accumulation for testing.
    
    \begin{figure}[H]
		\centering
		\includegraphics[width=0.9\textwidth]{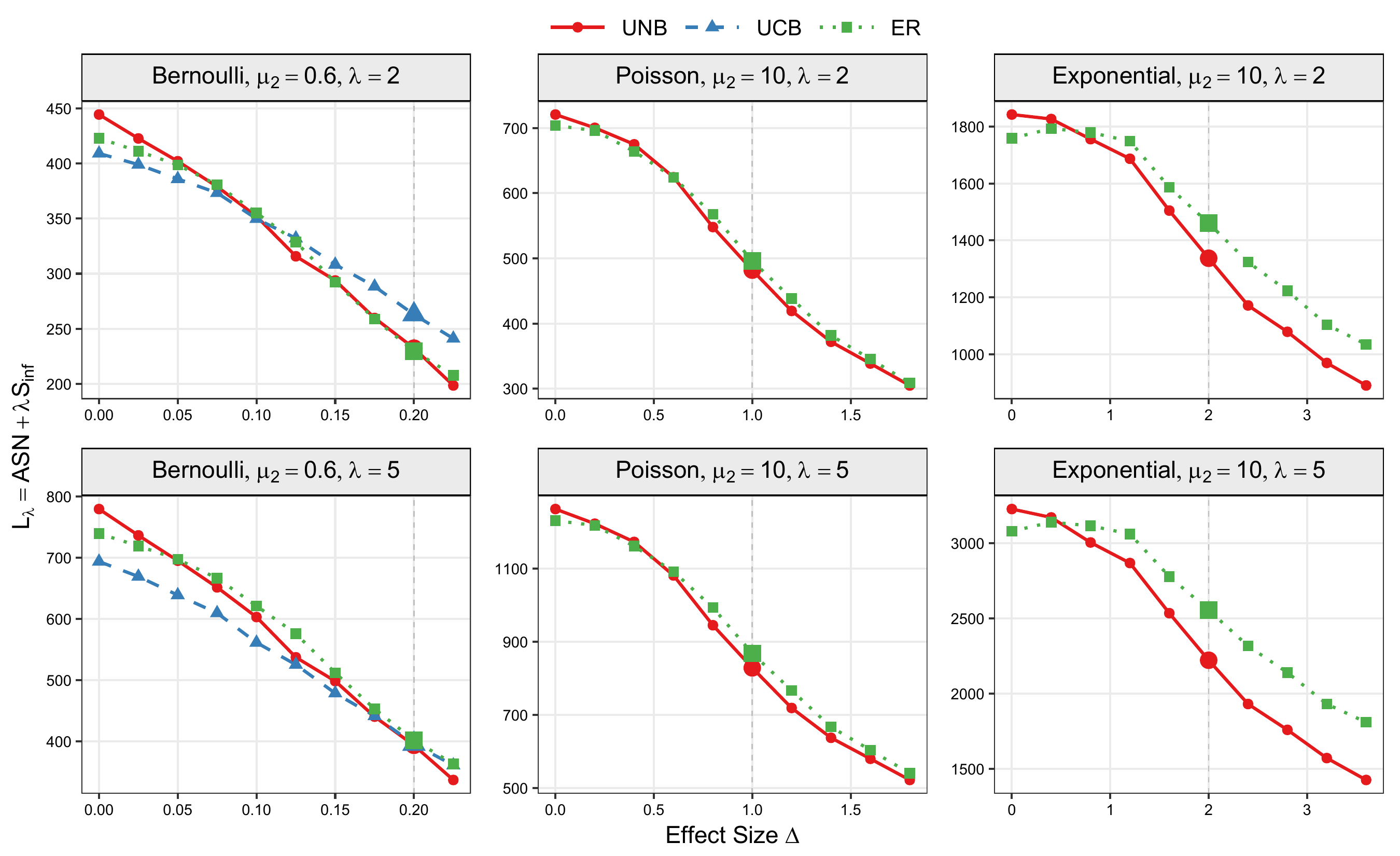}
		\vspace{-13pt}
		\caption{Loss index \eqref{eq:loss_function}, $L_{\lambda}=\mathrm{ASN}+\lambda S_{\text{inf}}$, 
        evaluated under different effect sizes $\Delta$ for the information-based sequential design with $\lambda=2$ (top) and $\lambda=5$ (bottom). The baseline means are set to $0.6$, $10$, and $10$ for the Bernoulli, Poisson, and Exponential rewards, respectively. The gray dashed vertical lines and enlarged markers denote the operating points corresponding to the pre-specified power level of $0.9$, at which UNB consistently attains smaller loss than both ER and UCB. The results for UCB under Poisson and Exponential rewards are omitted due to extreme Type I error inflation.}
		\label{figureloss}
	\end{figure}

    Ethics motivate allocating fewer observations to inferior arms, whereas inferential objectives favor allocations that preserve information for the target contrast (e.g., balanced sampling). 
    This ethical–statistical efficiency trade-off is well documented in response adaptive and bandit trial designs (see, e.g., \citet{Armitage1963, villar2015}). To summarize this trade-off on a single scale, we propose the weighted loss index:
	\begin{equation}\label{eq:loss_function}
		L_\lambda=\operatorname{ASN}+\lambda\,S_{\rm inf},
	\end{equation}
	where $\lambda>0$ encodes the relative emphasis placed on reducing inferior arm exposure versus reducing expected sample size. To evaluate how different allocation strategies balance this trade-off, we compare the weighted loss index of our proposed UNB with UCB and ER. In Figure~\ref{figureloss}, UNB matches or yields a smaller weighted loss $L_\lambda$ than ER and UCB across varying $\Delta$ for $\lambda \in \{2, 5\}$. Notably, the reduction in loss relative to ER is particularly pronounced under Exponential outcomes, reflecting a more effective trade-off between early stopping and inferior arm exposure.

	\section{Real Data Analysis}\label{sec6}
    This section evaluates our method using a real dataset, referred to as Dataset A, from a world leading ride-sharing company (anonymized for privacy). The data were collected through a randomized controlled trial in which users were randomly assigned to two different marketing strategies. To optimize economic return, the company aims to evaluate strategy performance while prioritizing better performing treatments and enabling early stopping once sufficient evidence is observed. Therefore, we formulate this as a sequential test problem: $H_0:\mu_1=\mu_2$ versus $H_1:\mu_1-\mu_2=\Delta>0$.

    To rigorously construct a valid null setting for evaluating empirical size, we first randomly permute the treatment assignments of the original data. This resampling procedure effectively neutralizes any unknown baseline differences and creates a valid A/A testing scenario. Subsequently, following common practice in methodological studies when proprietary experimental data are inaccessible \citep{li2023, luo2024, wen2025, zhang2025strategic}, we construct a semi-synthetic A/B testing dataset by injecting controlled treatment effects into the observed outcomes. This approach preserves the underlying distributional structure and heterogeneity of the real data while ensuring a controlled effect size for evaluation. The simulated results are presented as follows.

	\begin{figure}[H]
		\centering
		\includegraphics[width=0.8\textwidth]{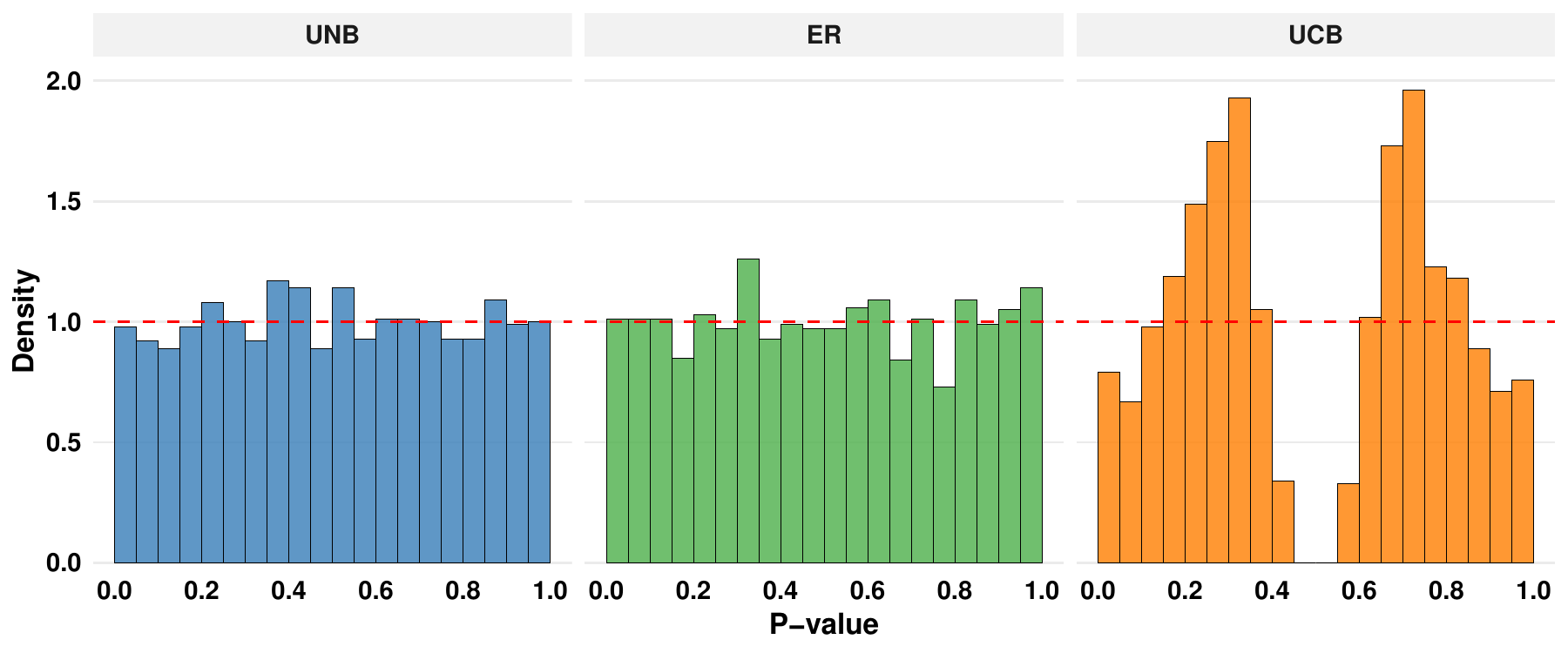}
		\vspace{-12pt}
		\caption{Empirical probability density of $p$-values under $H_0$ based on 2000 Monte Carlo samplings on the semi-synthetic real dataset. The red dashed line denotes the uniform distribution $U[0,1]$. Similar to the ER benchmark, UNB provides valid Type I error control.}
		\label{pvalue}
	\end{figure}

    Figure \ref{pvalue} shows that UNB and ER produce approximately Uniform$(0,1)$ distributed $p$-values under $H_0$, indicating valid Type I error control, whereas UCB exhibits substantial inflation due to bias induced by adaptive allocation. Consequently, UCB is excluded from subsequent evaluations of power and average reward lift.

	\begin{figure}[H]
		\centering
		\includegraphics[width=0.8\textwidth]{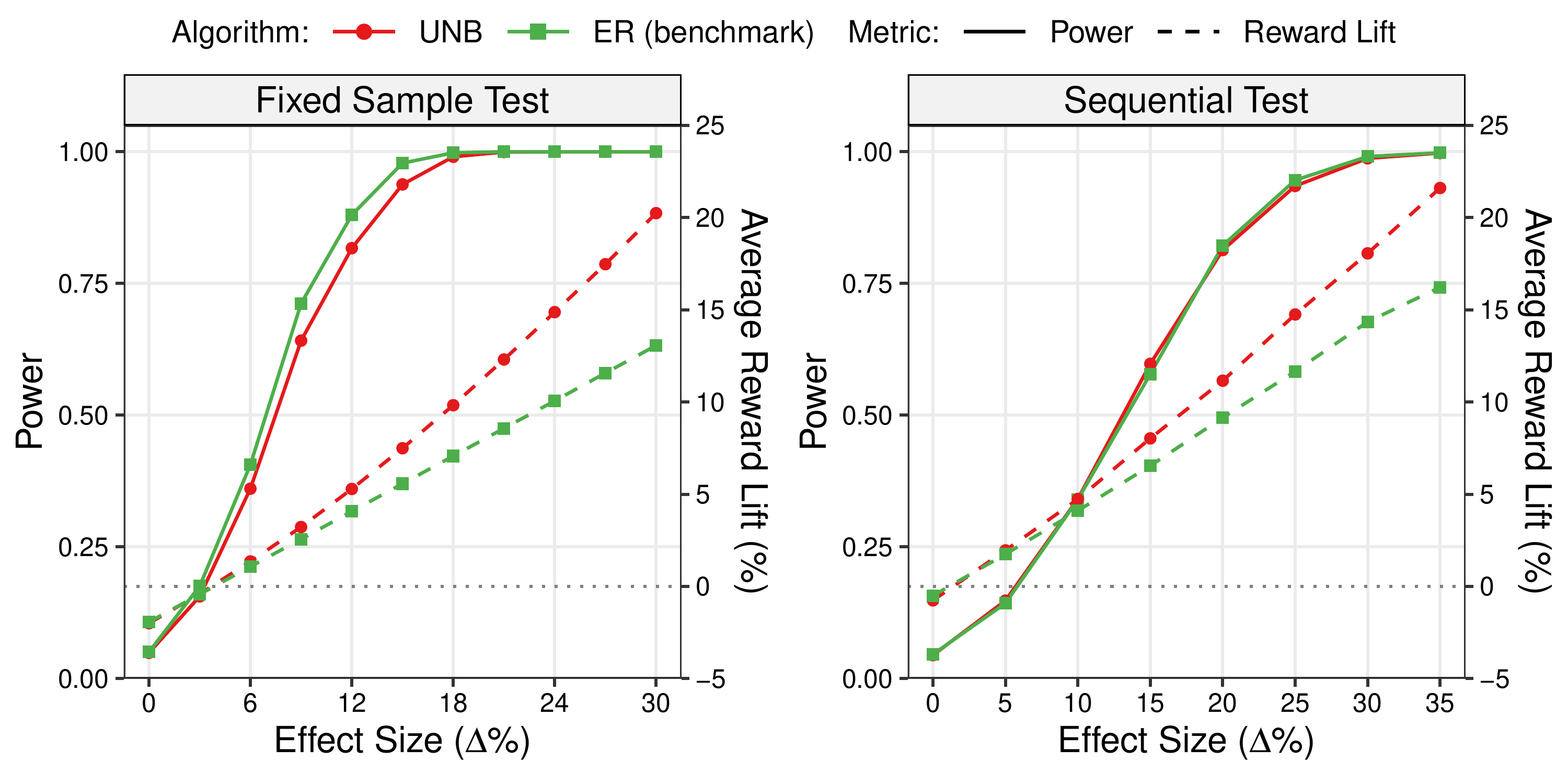}
		\vspace{-10pt}
		\caption{Performance comparison of allocation strategies under fixed sample tests (left) and sequential tests (right) frameworks. Solid lines (left y-axis) represent power of the test, while dashed lines (right y-axis) indicate the average reward lift (\%). UNB attains the benchmark power of ER but achieves a higher average reward lift.}
		\label{RealdataDouble}
	\end{figure}

    Figure \ref{RealdataDouble} (left panel) evaluates fixed-sample performance. UNB achieves power comparable to the ER benchmark while substantially improving efficiency by reducing inferior arm allocation (e.g., to $0.28$ at $\Delta=30\%$), yielding a $20.24\%$ reward lift relative to the pre-injection baseline. For sequential tests, we adopt an early-stopping policy where all remaining samples are assigned to the estimated optimal arm upon rejection. For each $\Delta$, reward lift is computed over a horizon defined by the larger average sample size of UNB and ER. As shown in Figure \ref{RealdataDouble} (right panel) and Table \tcc{S1} (in supplementary material), both methods often stop early, but UNB consistently outperforms ER due to adaptive allocation, yielding higher reward lifts (e.g., $11.16\%$ vs.  $9.15\%$ at 20\% $\Delta$, and $18.06\%$ vs.  $14.35\%$ at 30\% $\Delta$). This demonstrates that UNB yields a more favorable balance between statistical efficiency and overall reward.

	\section{Discussion}\label{sec7}
	This work advances the integration of adaptive allocation and sequential testing in experimental design and significantly improves both statistical efficiency and ethical performance of adaptive designs by dynamically prioritizing better performing treatments. Our theoretical results overcome the limitations of traditional methods, which often assume independent and sub-Gaussian observations and focus primarily on cumulative regret, thus enabling valid confidence intervals and hypothesis testing in complex scenarios. Further studies can focus on dynamic environments with nonstationary reward distributions, such as online recommendation systems with evolving user preferences. Computational complexity may pose challenges in large-scale applications with many arms.

		\begin{appendix}
		
		\section{Technical results}\label{appA}

		This section collects several key asymptotic properties of the UNB allocation process, which are foundational to our main results. We begin by establishing that each arm is sampled infinitely often and that its sample mean is strongly consistent.
       \begin{lemma}\label{ler1}
            Suppose that Assumptions \ref{meanbound} and \ref{meanconver_weak} hold. Then, for each arm $k\in[1:d]$, it holds almost surely that $S_{nk}\to\infty$ and $R_{nk}\to\infty$. Furthermore, 
        \begin{equation*}
            \hat\mu_{k,n}\xrightarrow{{\rm a.s.}} \mu_k.
        \end{equation*}
        \end{lemma}

        The following lemma establishes a strict lower bound on the growth rate of the total cumulative return, which governs the overall dynamics of the allocation process.
        \begin{lemma}\label{ler2}
            Suppose that Assumptions \ref{meanbound} and \ref{meanconver_weak} hold. For every $a<1$, it holds almost surely that $\liminf\limits_{n\to\infty}\frac{\|{\bf R}_n\|_1}{n}\ge a\mu_{\ast}>0$.
        \end{lemma}

        With the growth rate of the total reward bounded, we next investigate how the allocations are distributed among the arms. Lemma \ref{lerr4} establishes that the allocation proportions of suboptimal arms vanish asymptotically, while Lemma \ref{lerr1} establishes the asymptotic limits for the ratios of rewards and sample sizes between any pair of arms, revealing the asymptotic balance structure of the UNB algorithm. We recall that ${\mathcal{I}}=\{k:\mu_k = \mu^\ast\}$ denotes the set of optimal arms.
        
        \begin{lemma}\label{lerr4}
         Suppose that Assumptions \ref{meanbound} and \ref{meanconver_weak} hold. Then, it holds that
            $\sum\limits_{k\notin {\mathcal{I}}} Z_{nk} \xrightarrow{{\rm a.s.}} 0$.
    \end{lemma}

        \begin{lemma}\label{lerr1}
            Suppose that Assumptions \ref{meanbound} and \ref{meanconver_weak} hold. For all $k,j\in[1:d]$, there exists a random variable $\Lambda_{kj}$ on $(0,\infty)$ such that 
		      \begin{equation*}
                \frac{R^{1/\mu_k}_{nk}}{R^{1/\mu_j}_{nj}}\xrightarrow{{\rm a.s.}}\Lambda_{kj},\quad \frac{S^{1/\mu_k}_{nk}}{S^{1/\mu_j}_{nj}}\xrightarrow{{\rm a.s.}}\frac{\mu^{1/\mu_k}_k}{\mu^{1/\mu_j}_j}\Lambda_{kj}.
            \end{equation*}
        \end{lemma}

        To construct a valid plug-in covariance matrix for joint inference, it is imperative to estimate the cross-moments. Lemma \ref{ler5} guarantees this by proving that any arm pair with $\mu_k+\mu_s\ge\mu^\ast$ is drawn concurrently infinitely often.
       \begin{lemma}\label{ler5}
            Suppose that Assumptions \ref{meanbound} and \ref{meanconver_weak} hold. For $k,s$ such that $\mu_k+\mu_s\ge\mu^\ast$, it holds that $\sum\limits_{t=1}^n\mathbb{I}\{X_{tk}X_{ts}>0\}\to\infty$ almost surely when $N>1$.
        \end{lemma}

        Building on Lemma \ref{ler5}, the following result establishes the strong consistency of moment and cross-moment estimators, enabling the construction of $\widehat{\bm{\Sigma}}_n$.
        \begin{lemma}[Parameter estimation]\label{lerr2}
		Suppose that Assumptions \ref{meanbound} and \ref{meanconver} hold. Let $q_{k,n}=\mathbb{E}(\xi^2_{nk,q})\to q_k$ and $q_{ks,n}=\mathbb{E}(\xi_{nk,q}\xi_{ns,q})\to q_{ks}$.
		For all $k,s\in[1:d]$, the following holds almost surely{\rm :}
		\begin{enumerate}[label=\textup{(\arabic*)}]
			\item Return Proportions{\rm :} $\hat{Z}_{k,n}=\frac{1}{n}\sum_{j=1}^n\frac{X_{jk}}{N_j} \to Z_k$.
				
			\item Moment Estimators{\rm :} The moment estimator
			\begin{equation*}
				\hat{q}_{k,n}=\frac{\sum_{t=1}^n\sum_{q=1}^{X_{tk}}\xi^2_{tk,q}}{S_{nk}}
             \end{equation*}
            converges almost surely to $q_k$. Moreover, for any $k \neq s$ such that $\mu_k+\mu_s\ge\mu^\ast$, the cross-moment estimator
            \begin{equation*}
                \hat{q}_{ks,n}=\frac{\sum_{t=1}^n\sum_{q=1}^{X_{tk}\wedge X_{ts}}\xi_{tk,q}\xi_{ts,q}}{\sum_{t=1}^n X_{tk}\wedge X_{ts}}
			\end{equation*}
           converges almost surely to $q_{ks}$ when $N>1$, where $X_{tk}\wedge X_{ts}:=\min\{X_{tk},X_{ts}\}$.
		\end{enumerate}
        \end{lemma}

		Based on the estimators of the second moments and cross moments of $\{\xi_{n,k}\}$ provided in Lemma \ref{lerr2}, the estimators of the cross-arm covariance and correlation coefficients are given by 
			$\hat{C}_{ks,n}=\hat{q}_{ks,n}-\hat{\mu}_{k,n}\hat{\mu}_{s,n}$ and $\hat{\rho}_{ks,n}=\frac{\hat{C}_{ks,n}}{{\hat{\sigma
            }_{k,n}\hat{\sigma}_{s,n}}}$, where $\hat{\sigma}^2_{k,n}=\hat{q}_{k,n}-\hat{\mu}^2_{k,n}$.

        By Lemma \ref{lerr2}, if $\sum_{t=1}^n\mathbb{I}(X_{tk}X_{ts}>0)\xrightarrow{{\rm a.s.}}\infty$, then $\hat{q}_{ks,n}\xrightarrow{{\rm a.s.}} q_{ks}$ when $N>1$. On the event $\{\sum_{t=1}^\infty \mathbb{I}(X_{tk}X_{ts}>0)=0\}$, we define $\hat q_{ks,n}=0$. We next consider the complementary case where $0<\sum_{t=1}^\infty\mathbb{I}(X_{tk}X_{ts}>0)<\infty$, and establish a boundedness property of $\hat{q}_{ks,n}$.
            \begin{lemma}\label{ler3}
                Suppose that Assumptions \ref{meanbound} and \ref{meanconver_weak} hold. On the event $\{0<\sum_{t=1}^\infty\mathbb{I}(X_{tk}X_{ts}>0)<\infty\}$, it holds almost surely that
            \begin{equation*}
                \limsup_{n\to\infty} \hat{q}_{ks,n}=\limsup_{n\to\infty} \frac{\sum_{t=1}^n\sum_{q=1}^{X_{tk}\wedge X_{ts}}\xi_{tk,q}\xi_{ts,q}}{\sum_{t=1}^n X_{tk}\wedge X_{ts}}<\infty.
            \end{equation*}
            \end{lemma}
	\end{appendix}


\bibliographystyle{apalike} 

\bibliography{reference}

\end{document}